\documentclass{ieeeaccess}

\usepackage[utf8]{inputenc}
\usepackage[T1]{fontenc}
\usepackage{inconsolata}
\usepackage[english]{babel}

\usepackage{amsmath,amssymb,amsfonts}
\usepackage{algorithmic}
\usepackage{graphicx}
\usepackage{textcomp}

\usepackage{tikz}
\usetikzlibrary{shapes.geometric, arrows}

\usepackage{bm}
\makeatletter
\AtBeginDocument{\DeclareMathVersion{bold}
\SetSymbolFont{operators}{bold}{T1}{times}{b}{n}
\SetSymbolFont{NewLetters}{bold}{T1}{times}{b}{it}
\SetMathAlphabet{\mathrm}{bold}{T1}{times}{b}{n}
\SetMathAlphabet{\mathit}{bold}{T1}{times}{b}{it}
\SetMathAlphabet{\mathbf}{bold}{T1}{times}{b}{n}
\SetMathAlphabet{\mathtt}{bold}{OT1}{pcr}{b}{n}
\SetSymbolFont{symbols}{bold}{OMS}{cmsy}{b}{n}
\renewcommand\boldmath{\@nomath\boldmath\mathversion{bold}}}
\makeatother

\def\BibTeX{{\rm B\kern-.05em{\sc i\kern-.025em b}\kern-.08em
    T\kern-.1667em\lower.7ex\hbox{E}\kern-.125emX}}

\usepackage{acronym}
\usepackage{csquotes}

\usepackage{listings}
\usepackage{xcolor}

\definecolor{IEEEblue}{rgb}{0.0, 0.0, 0.5}
\definecolor{IEEElightblue}{rgb}{0.0, 0.4, 0.8}
\definecolor{IEEEgray}{rgb}{0.4, 0.4, 0.4}

\usepackage[
    backend=biber,
    style=numeric,
    sorting=none
]{biblatex}
\usepackage{pgfplots}
\usepgfplotslibrary{groupplots}
\pgfplotsset{
    table/search path={data},
    compat=1.18
}

\usepackage{multirow}
\usepackage{makecell}
\usepackage{colortbl} 
\usepackage{siunitx}

\usepackage{hyperref}

\graphicspath{ {./images/} }

\definecolor{lcolor}{HTML}{5B72E6}
\definecolor{mcolor}{HTML}{E2EA76}
\definecolor{hcolor}{HTML}{EF7458}

\NewSpotColorSpace{PANTONE}
\AddSpotColor{PANTONE}{PANTONE3015C}{PANTONE\SpotSpace 3015\SpotSpace C}{1 0.3 0 0.2}
\SetPageColorSpace{PANTONE}

\definecolor{accessblue}{HTML}{1976ac}
\definecolor{greycolor}{HTML}{58585a}

\newcolumntype{u}{>{\columncolor{gray!15}} c}
\newcolumntype{v}{>{\columncolor{gray!30}} c}
\newcolumntype{q}{>{\columncolor{gray!15}} l}

\newcommand{\markyes}{\checkmark}

\newcommand{\papertitle}{Metrics for Assessing Changes in Flow-based Networks}
\newcommand{\mainauthorname}{Michał Rzepka}
\newcommand{\mainauthormail}{michal.rzepka@outlook.com}
\newcommand{\secauthorname}{Piotr Chołda}
\newcommand{\secauthormail}{piotr.cholda@agh.edu.pl}

\begin{document}

\acrodef{DTB}[DTB]{Delay-Tolerant Bulk}
\acrodef{SDN}[SDN]{Software-Defined Networking}
\acrodef{GFU}[GFU]{Generalized Flow Utilization}
\acrodef{SNMP}[SNMP]{Simple Network Management Protocol}
\acrodef{ISP}[ISP]{Internet Service Provider}
\acrodefplural{ISPs}[ISPs]{Internet Service Providers}
\acrodef{IPFIX}[IPFIX]{Internet Protocol Flow Information Export}
\acrodef{QoS}[QoS]{Quality of Service}
\acrodef{OWAMP}[OWAMP]{One-Way Active Management Protocol}
\acrodef{TWAMP}[TWAMP]{Two-Way Active Management Protocol}
\acrodef{AS}[AS]{Autonomous System}
\acrodef{TCP}[TCP]{Transmission Control Protocol}
\acrodef{UDP}[UDP]{User Datagram Protocol}
\acrodef{IPv4}[IPv4]{Internet Protocol Version 4}
\acrodef{IPv6}[IPv6]{Internet Protocol Version 6}
\acrodef{bps}[bps]{bits per second}
\acrodef{BRIAN}[BRIAN]{Backbone Router Interface Analytics}
\acrodef{API}[API]{Application Programming Interface}
\acrodef{LP}[LP]{Linear Programming}
\acrodef{ABW}[ABW]{Available Bandwidth}
\acrodef{TE}[TE]{Traffic Engineering}
\acrodef{LUPD}[LUPD]{Link Utilization Percentile Delta}
\acrodef{TLUSSD}[TLUSSD]{Top Link Utilization Samples Share Delta}
\acrodef{LUSD}[LUSD]{Link Utilization Score Delta}
\acrodef{MLUPD}[MLUPD]{Mean Link Utilization Percentile Delta}
\acrodef{TMLUSSD}[TMLUSSD]{Top Mean Link Utilization Samples Share Delta}
\acrodef{GGPD}[GGPD]{Growth GFU Percentile Delta}
\acrodef{RGPD}[RGPD]{Risk GFU Percentile Delta}
\acrodef{TGGSSD}[TGGSSD]{Top Growth GFU Samples Share Delta}
\acrodef{TRGSSD}[TRGSSD]{Top Risk GFU Samples Share Delta}
\acrodef{LUPSV}[LUPSV]{Link Utilization Percentile Shapley Value}
\acrodef{TLUSSSV}[TLUSSSV]{Top Link Utilization Samples Share Shapley Value}
\acrodef{GÉANT}{Gigabit European Academic Network}

\history{-}
\doi{-}

\title{\papertitle}
\author{\MakeUppercase{\mainauthorname}\authorrefmark{1}, \MakeUppercase{\secauthorname}\authorrefmark{2}}
\address[1]{Independent researcher, Kraków, Poland ({e-mail: \mainauthormail})}
\address[2]{Institute of Telecommunications, AGH University, Kraków, Poland ({e-mail: \secauthormail})}

\tfootnote{}

\markboth
{\mainauthorname,~\secauthorname: \papertitle}
{\mainauthorname,~\secauthorname: \papertitle}

\corresp{Corresponding author: \secauthorname~(e-mail: \secauthormail).}

\begin{abstract}
This paper addresses the challenges of evaluating network performance in the presence of fluctuating traffic patterns, with a particular focus on the impact of peak data rates on network resources. We introduce a set of metrics to quantify network load and measure the impact of individual flows on the overall network state. By analyzing link and flow data through percentile values and sample distributions, and introducing the Utilization Score metric, the research provides insights into resource utilization under varying network conditions. Furthermore, we employ a modified Shapley value-based approach to measure the influence of individual flows on the network, offering a better understanding of their contribution to network performance. The paper reviews and compares 11 metrics across various network scenarios, evaluating their practical relevance for research and development. Our evaluation demonstrates that these metrics effectively capture changes in network state induced by specific flows, with three of them offering a broad range of valuable insights while remaining relatively easy to maintain. Moreover, the methodology described in this paper serves as a framework for future research, with the potential to expand and refine the set of metrics used to evaluate flow impact on network performance.
\end{abstract}


\titlepgskip=-21pt

\maketitle

\section{Introduction}

\PARstart{E}{nsuring} network operability despite unexpected traffic fluctuations is one of the main goals of network management~\cite{tizghadam:congestion}. Network resource performance is directly related to link utilization across the topology. Although past research suggested that backbone links were often underutilized~\cite{hassidim:gfu}, recent deployments show links reaching 100\% utilization~\cite{rzym:dtm}. This necessitates thorough network planning and monitoring to maintain adequate \ac{QoS}, as overutilized links cause delays, packet losses, and reduced traffic handling capability. Such situations should be promptly detected, thoroughly investigated, and, preferably, avoided at all times. Therefore, a thorough traffic analysis is necessary to accurately identify contention periods and pinpoint flows that may contribute to congestion. 

Beyond monitoring, network services can be designed to reduce traffic peaks, preventing resource overutilization and network path congestion. Examples include scheduling of \ac{DTB} data transfers to off-peak hours~\cite{laoutaris:dtb} and shifting the timing of polling messages in an OpenFlow-based \ac{SDN} environment~\cite{rzepka:sfsa}. Such solutions successfully reduce the maximum transient data rate while ensuring complete data transfer and meeting other applicable requirements~\cite{zining:percentile-optimization}.

Effective network monitoring and research on resource-efficient mechanisms require appropriate metrics and analytical methods to accurately identify congestion, assess flow impact, and evaluate countermeasure effectiveness. However, analyzing the network load at the flow level is not trivial. Traffic demands change dynamically over time and space~\cite{hassidim:gfu}. Network flows are generated by various users or services, each serving a distinct purpose. These flows inherently exhibit differences in their traffic characteristics~\cite{oudah:burstiness}. Even when transferring the same volume of data over a given period, some flows maintain a stable data rate, while others generate transient traffic bursts that may temporarily saturate network links.

Therefore, when assessing the impact of these flows on the network, it is important to note that merely comparing flow rates over a specified period is insufficient~\cite{koumar:classification}. What truly matters is their contribution to link congestion, their potential to restrict other flows from accessing network resources~\cite{briscoe:internet, briscoe:fairness}, and their impact on \ac{QoS} degradation~\cite{islam:harm}. Given these considerations, a comprehensive analysis of available solutions is required to align use cases with appropriate metrics.

In this study, we aim to discuss, compare, and propose metrics for evaluating the impact of a flow on overall network load. The ideal metric should capture key characteristics of network traffic related to congestion or peak data rates and assist in determining whether a particular flow is likely to cause resource starvation. We explore mechanisms that achieve the following objectives:
\begin{enumerate}
    \item Collecting measurement data,
    \item Assessing network load based on peak data rates,
    \item Quantifying the impact of specific flows on network state.
\end{enumerate}
The ultimate objective of this research is to develop simple and interpretable metrics that effectively characterize network changes and facilitate the analysis of collected measurement data.

To provide insight into existing solutions, we examine measurement techniques, data sources, and the most significant features of solutions discussed in the related literature, including their use cases and potential research directions. Several metrics (e.g., link utilization, \ac{GFU}) and analysis methods (e.g., percentiles, sample distribution properties, Shapley value) are considered. Additionally, as part of our contribution, we evaluate some techniques already employed for other purposes, such as billing, for their applicability to traffic engineering and benchmarking. We extend existing concepts from related work and introduce new metrics.

First, we propose a general taxonomy for designing such metrics. This approach defines a four-tier metric hierarchy, with \emph{network load metrics} and \emph{flow impact metrics} as the primary focus of this research. We examine the application of percentile analysis to time series of link utilization and \ac{GFU}. In addition, we introduce a \emph{network load metric} called the Utilization Score. The proposed \emph{network load metrics} offer a means to assess network resource utilization. Second, we modify the Shapley value-based analysis method and analyze differences in metric values to develop \emph{flow impact metrics}. Finally, we evaluate the defined metrics across various network scenarios using dedicated datasets. Based on our observations, we draw conclusions about the usability of each metric. The presented overview of the evaluated solutions aims to support the selection of benchmarking methods for researching and developing network mechanisms and assessing the importance of traffic components. 

The remainder of the paper is structured as follows. Section~\ref{sec:background} provides background on previously proposed metrics in related research. Additionally, we introduce the assumptions underlying the selection of metrics for this research. Sections~\ref{sec:basic-metrics} and~\ref{sec:load-impact-metrics} describe all metrics examined in this paper, including their key features and implementation details. Section~\ref{sec:experiments} presents the experiments conducted to evaluate the effectiveness of the metrics, along with the results obtained throughout the study. Section~\ref{sec:conclusions} provides the conclusions of this study.

\section{Background and related work}
\label{sec:background}

This section provides an overview of traffic measurement techniques, discusses relevant research on network load evaluation and flow impact, and presents our rationale for selecting specific solutions for examination.

The acquisition of measurement data is fundamental to network monitoring and traffic analysis. Collecting parameters related to network links and nodes is the first step in assessing the state of network infrastructure and its services, irrespective of the analysis methods employed. Measurement methods are generally classified as active and passive~\cite{tan:int-survey}.

Active monitoring methods provide a means of understanding network performance and assessing network load. Throughput, jitter, packet loss, and latency are a few \ac{QoS} parameters that offer valuable insights into traffic handling efficiency~\cite{he:measurements-survey}. This approach requires generating traffic using tools ranging from simple utilities (\emph{ping}, \emph{iperf}, \emph{traceroute}) to more complex implementations (RFC~6349, \ac{OWAMP}, \ac{TWAMP}), potentially introducing disruptive overhead to control or data planes~\cite{tahaei:sdn-measurement}. 

In~\cite{yi:performance}, the method of obtaining an individual link latency is followed by steps to normalize the collected data and derive an evaluation coefficient. The coefficient is then calculated at 3 different levels of granularity: link, city and \ac{AS}. Its value is intended to represent overall network load and performance. This method presents a simplified approach compared to studies that utilize bandwidth and data loss for metric calculation.

One such solution supplements active delay measurements with throughput monitoring~\cite{narayan:performance}. The performance of network infrastructure running an \ac{IPv4}/\ac{IPv6} stack is evaluated with regard to the number of routers and packet size. The study investigates the impact of these factors on \ac{TCP} and \ac{UDP} traffic. Through the conducted experiments, the researchers observe a relationship between network configuration, selected protocols, and \ac{QoS} parameters. However, their research considers only average and maximum metric values.

Conversely, network infrastructure offers straightforward methods for passive traffic monitoring. The volume of data transferred through each interface can be obtained by periodically polling designated packet or byte counters. These mechanisms can be implemented using \ac{SNMP} or OpenFlow. More advanced solutions include sFlow, NetFlow, and \ac{IPFIX}. Since these methods do not generate probe traffic, they are considered non-intrusive~\cite{tahaei:sdn-measurement}. However, running these mechanisms may still impact the control plane~\cite{megyesi:bandwidth}. These methods provide data rates for interfaces and even selected flows. Interface rate, combined with link capacity, can be used to calculate link utilization percentage. 

Based on these findings, in this paper we focus on solutions that utilize information obtained by passive measurements. The rationale for selecting these techniques is that they rely on data that is likely already being collected and generally do not require intrusive modifications to network monitoring mechanisms. Analysis can be performed on an arbitrary set of historical traffic statistics using external tools. Notably, the underlying concepts of billing mechanisms suitable for \acp{ISP} closely align with the objectives of this research. Both aim to quantify the effort required to operate the network or process traffic from specific flows. Therefore, metrics based on passively collected traffic data are initially taken into consideration.

Aggregating traffic volume time series in 5-minute bins and calculating the 95$^{th}$ percentile is an industry standard for transit billing used by \acp{ISP}~\cite{zhan:cost-aware-te, huawei:cloud-billing, akamai:cloud-billing}. The resulting value estimates the network load generated by the customer while disregarding the top 5\% of traffic bursts. With an appropriate model, this scheme can also be used to forecast future bills~\cite{alasmar:log-normal}. Intuitively, this metric is expected to represent the predominant saturation level of the network or how it is impacted by customer traffic. Due to its popularity, the method is thoroughly discussed in the research. Despite its ease of use, several flaws have also been identified. The authors of~\cite{dimitropoulos:percentile} found that the choice of aggregation window size significantly (up to approx.~40\%) impacts the results. However, it is not possible to find a definitive set of parameters for an ideal metric. Instead, they propose standardizing the window size across all network operators. Another study~\cite{vamseedhar:percentile} questions the fairness of the percentile method. According to a comprehensive analysis, while easy to calculate, the 95$^{th}$-percentile method does not accurately reflect the provider's perceived cost burden of handling customer traffic. As a result, the percentile-based method may not be the sole appropriate approach for assessing the impact of flows on the network. These findings justify exploring alternative solutions to capture contributions to overall network load.

Consequently, the Shapley Value Percentile Billing method is examined~\cite{stanojevic:shapley}. This concept, derived from game theory, models the percentile value of network traffic as the cost function of a cooperative game. By evaluating the average marginal contribution of each traffic component, this method balances aggregate usage and peak usage. The Shapley value method effectively represents a user's contribution to total network load at the cost of increased computational complexity. However, the computational load can be reduced by utilizing the Monte Carlo approximation method. As a result, the authors of~\cite{stanojevic:shapley} conclude that the~Shapley value provides a promising general framework for fair cost-sharing and can be applied to various metrics measuring peak traffic demand. Additionally, a Shapley value-based method has been proposed to allocate network investment and maintenance costs to residential customers~\cite{azuatalam:shapley-pricing,azuatalam:turvey-shapley}. These findings support the consideration of Shapley value-based metrics for assessing the impact of flows on the network.

Unlike billing mechanisms that analyze traffic volume using various methods, the \ac{ABW} metric offers an alternative approach to characterizing a network~\cite{megyesi:bandwidth}. It represents the average unused capacity on a link over a given time interval. Although the original study applies this metric to individual network paths, the method could be extended to assess the entire network. The results of individual measurements could then be compared to identify differences caused by specific flows. However, unused capacity is closely tied to link utilization, a metric already considered in our research, and is therefore unlikely to offer significant additional insights in the current research.

While the aforementioned metrics focus on link utilization data, an alternative concept of a flow-based view is also proposed in related work~\cite{hassidim:gfu}. Since link utilization does not necessarily reflect network performance or user experience, a new measure, \acf{GFU}, is introduced. The \ac{GFU} is meant to serve as a general framework for evaluating the network’s ability to support current demands and potential changes. Its versatility stems from its capability to incorporate various utilization functions, such as those addressing flow growth accommodation. Therefore, examining this metric, particularly its variant focused on flow growth, could be valuable for assessing network state and understanding how it is affected by specific flows. The original research paper suggests that \ac{GFU} can effectively support network planning and identify traffic at risk. However, computing the required $\alpha^{growth}$ vector is resource-intensice and time-consuming, as is obtaining a required list of all flows present in the network.

Beyond specific measurement techniques, \cite{zhang:evaluation} introduces two methods for analyzing measurement metrics. These methods capture relative changes in metric values across consecutive time windows. Additionally, the study provides an algorithm for removing outlier samples from datasets. The resulting solutions enable the detection of performance degradation caused by increased network load during infrastructure outages, attacks, or other anomalies related to exceptionally high resource usage. Although these techniques are beyond the scope of this paper, they could inform future research on extending the developed metrics.

In conclusion, related research offers a variety of concepts for developing tools to evaluate network load and assess the impact of specific traffic components on the network. The methods discussed above are primarily related to active or passive measurements and adopt a link- or flow-oriented perspective of the network. The last of the cited studies introduces a metric-agnostic analytical method, making a versatile contribution to the state-of-the-art in this field. However, certain technique combinations and application areas remain unexplored. Existing research lacks a comprehensive overview of available solutions and their characteristics. Furthermore, the efficiency of these methods in benchmarking network mechanisms remains insufficiently evaluated. Therefore, in this paper we analyze these techniques to address the existing research gap. Our study compares both original proposals and existing methods to support the selection of tools for processing measurement data.

\section{Basic metrics and techniques}
\label{sec:basic-metrics}

This section aims to systematize the researched concepts and provide a detailed explanation of the approach used to define the metrics. First, we introduce general concepts related to measurement data analysis. Next, we present the desired features of the metrics. The remainder of the section lists known metrics and analysis methods along with their characteristics. The section concludes with a summary of the discussed techniques.

\subsection{General concepts}

\begin{figure}[!ht]
    \centering
    \includegraphics[width=\columnwidth]{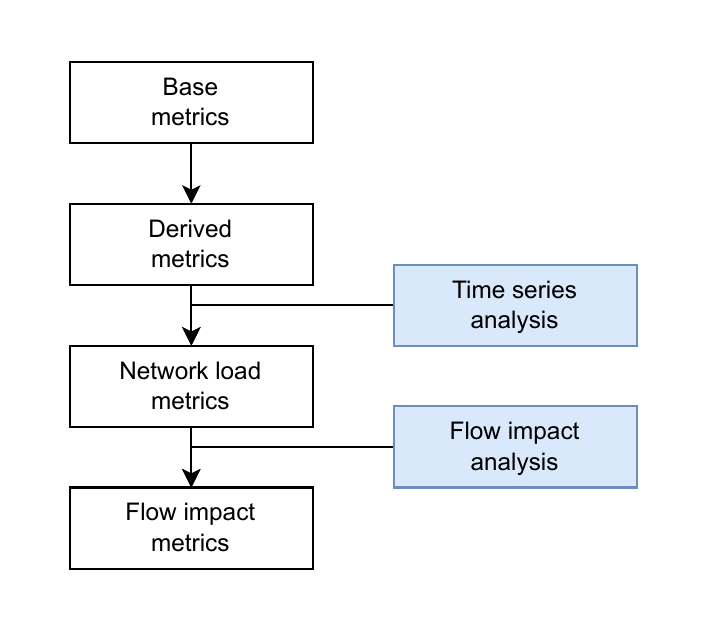}
    \caption{Metric hierarchy and stages of analysis}
    \label{fig:metric-hierarchy}
\end{figure}

As mentioned above, this paper aims to evaluate methods for assessing the impact of specific flows on network load. Each discussed approach is based on fundamental measures used to characterize link and flow properties within the network. However, certain steps must be taken to process the initial data and achieve the stated objectives. These steps include normalization, generalization, and other dataset analysis techniques (Fig.~\ref{fig:metric-hierarchy}). To effectively represent this process, we define a four-tier metric hierarchy:
\begin{enumerate}
    \item base metrics,
    \item derived metrics,
    \item network load metrics,
    \item flow impact metrics.
\end{enumerate}
The first two tiers primarily rely on concepts established in related research. The primary contribution of this paper is the design of the $3^{rd}$- and $4^{th}$-tier metrics, which directly address challenges in describing network load and flow contributions. The role of each tier in this process is described in the following paragraphs.

\begin{figure}[t]
  \begin{center}
    \begin{tikzpicture}
      \begin{axis}[        
        table/col sep=comma,
        width=\columnwidth,
        height=0.5\columnwidth,
        xlabel={Time},
        grid=major,
        xmajorticks=false,
        xminorticks=false,
        ymin=0,ymax=1.1e11,
        ylabel={Traffic throughput [bps]},
        legend entries={Link \#1, Link \#2},
        legend style={anchor=center, at={(0.85,0.45)}}
      ]
        \addplot+[red,mark size=0.1pt] table[x=index,y=bw-ams-fra] {link_measurements_example.csv};
        \addplot+[blue,mark size=0.1pt] table[x=index,y=bw-fra-poz]
        {link_measurements_example.csv};
      
      \end{axis}
    \end{tikzpicture}
    \caption{Time series of throughput}
    \label{fig:example-throughput-time-series}
  \end{center}
\end{figure}
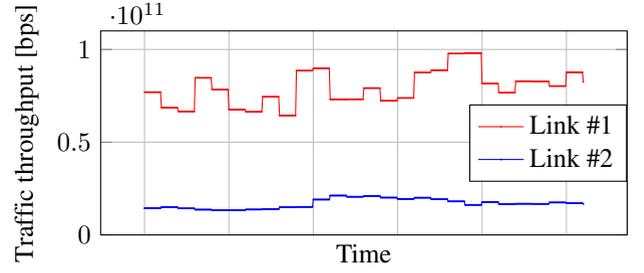

The first tier consists of the simplest metrics periodically collected during network monitoring. This paper primarily examines throughput, which, alongside delay and packet loss, constitutes the \emph{base metrics} described in Section~\ref{subsec:base-metrics}. Datasets generated by collecting individual metric samples over a specific time period and at various locations across the network illustrate changes in particular links, nodes, or flows. Such time series for two network links are presented in Fig.~\ref{fig:example-throughput-time-series}.

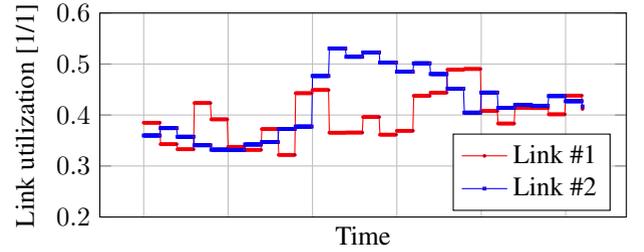
\begin{figure}[t]
  \begin{center}
    \begin{tikzpicture}
      \begin{axis}[        
        table/col sep=comma,
        group style={
            group size=2 by 1,
            xlabels at=edge bottom
        },
        width=\columnwidth,
        height=0.5\columnwidth,
        xlabel=Time,
        grid=major,
        xmajorticks=false,
        xminorticks=false,
        ymin=0.2,ymax=0.6,
        ylabel={Link utilization [1/1]},
        legend entries={Link \#1, Link \#2},
        legend pos=south east
      ]
      
        \addplot+[red,mark size=0.5pt] table[x=index,y=utilization-ams-fra] {link_measurements_example.csv};
        \addplot+[blue,mark size=0.5pt] table[x=index,y=utilization-fra-poz] {link_measurements_example.csv};
      
      \end{axis}
    \end{tikzpicture}
    \caption{Time series of link utilization}
    \label{fig:example-link-utilization-time-series}
  \end{center}
\end{figure}

\begin{figure}[t]
  \begin{center}
    \begin{tikzpicture}
      \begin{axis}[
          table/col sep=comma,
          height=0.5\columnwidth,
          width=\columnwidth,
          grid=major,
          legend entries={Mean utilization, Growth GFU},
          xlabel=Time,
          ylabel={Metric value [1/1]},
          ymin=0, ymax=0.4,
          xmajorticks=false,
          xminorticks=false
        ]
        
        \addplot+[red,dashed,mark size=0.5pt]
            table[x expr=\coordindex,y=mean-utilization] {data_series_example.csv}; 
        \addplot+[blue,dashed,mark size=0.5pt]
            table[x expr=\coordindex,y=growth-gfu] {data_series_example.csv};
            
      \end{axis}
    \end{tikzpicture}
    \caption{Time series of \ac{GFU} and mean link utilization}
    \label{fig:example-network-time-series}
  \end{center}
\end{figure}
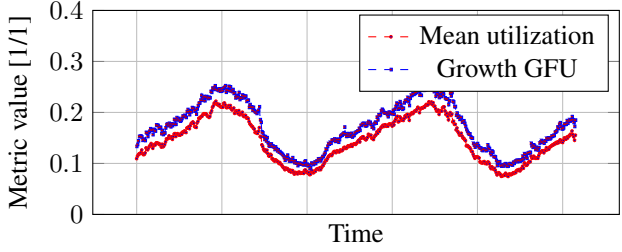

Building on these metrics, more advanced ones can be derived. Traffic throughput divided by the corresponding link capacity yields link utilization values (Fig.~\ref{fig:example-link-utilization-time-series}). The mean link utilization across all links at a given timestamp serves as another related metric. Furthermore, knowledge of network flows and link utilization values enables the calculation of \ac{GFU}. Example values of both mean link utilization and \ac{GFU} are shown in Fig.~\ref{fig:example-network-time-series}. These metrics, referred to as \emph{derived metrics} in this paper, are described in Section~\ref{subsec:derived-metrics}.

\begin{figure}[t]
  \begin{center}
    \begin{tikzpicture}
        \begin{groupplot}[        
            table/col sep=comma,
            group style={
                group size=2 by 1,
                xlabels at=edge bottom,
                horizontal sep=1.2cm
            },
            width=0.5\columnwidth,
            height=0.75\columnwidth,
            xlabel={Samples [\#]},
           xmajorticks=false,
            xminorticks=false,
            grid=major,
            scaled x ticks=false,
            x tick label style={/pgf/number format/fixed}
        ]

        \nextgroupplot[ymin=0,ymax=1.1e11,legend entries={Throughput},ylabel={Traffic throughput [bps]}]
        \addplot+[mark size=0.5pt] table[x=index,y=throughput] {distribution_example.csv};

        \nextgroupplot[ymin=0,ymax=1,legend entries={Utilization},ylabel={Link utilization [1/1]}]
        \addplot+[mark size=0.5pt] table[x=index,y=utilization] {distribution_example.csv};

        \end{groupplot}
    \end{tikzpicture}
    \caption{Distributions of throughput and link utilization samples}
    \label{fig:example-distribution}
  \end{center}
\end{figure}
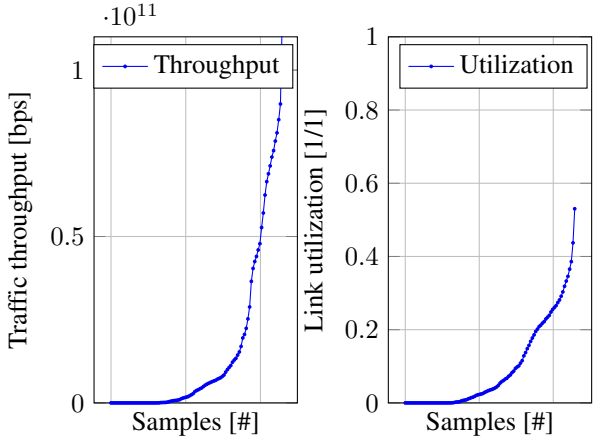

Although the aforementioned metrics effectively capture specific aspects of network status at a given timestamp, their time series are challenging to classify as a whole. A desirable outcome of further analysis would be the assignment of a single value summarizing network status over a specified time window. A possible solution involves analyzing the distribution of all dataset samples (Fig.~\ref{fig:example-distribution}) and extracting information about its characteristics~\cite{koumar:classification}. Therefore, the $3^{rd}$-tier metrics proposed in this paper employ appropriate analysis methods (e.g., percentile calculation and others listed in Section~\ref{subsec:ts-analysis}) to derive a single value representing the entire dataset. In this paper, these metrics are referred to as \emph{network load metrics} and are presented in Section~\ref{subsec:network-load-metrics}.

The $3^{rd}$-tier metrics can be used to provide a straightforward description of network state over a selected time period. However, determining how a given network state is influenced by a specific flow requires further analysis. This purpose is fulfilled by $4^{th}$-tier \emph{flow impact metrics}, which integrate lower-tier metrics with additional analysis methods. These metrics quantify the contribution of an examined traffic component to the overall metrics representing network state. To achieve this, Shapley value analysis or a simple comparison of metric values may be applied (cf. Section~\ref{subsec:contrib-analysis}). The \emph{flow impact metrics} constitute the final contribution of this paper. 

In summary, \emph{base metric} samples are collected over a period of time, normalized, and used to calculate more complex \emph{derived metrics}. To generalize the resulting time series, appropriate analysis methods are applied. The~\emph{network load metrics} yield a single numerical value representing the entire processed dataset. This value is then used to characterize network state over the selected time period. Finally, the impact of certain flows on \emph{network load metrics} can be determined using \emph{flow impact metrics}. The proposed metric hierarchy and the process of formulating the metrics function as a general framework that can be extended through future research with new metrics and analysis methods.

When describing the metrics, we use the following notation:

\begin{itemize}
    \item $G = (V,E)$ is a graph that represents the network topology, where $|E|$ is the number of network links with a corresponding capacity vector $C$, and $|V|$ the number of network nodes.
    \item It is assumed that monitoring the network over a $t$ period of time yields $w = t/i$ samples for each monitored location $l = 0, 1, ..., n$ (link or flow), where $i$ denotes the measurement interval. The complete set of collected samples is denoted by $M$ and can be stored as a matrix of size $w \times n$. The matrix consists of vectors containing sequential values measured at specific locations. The measurement value at location $l$ at timestamp $w$ is denoted as $M_{l,w}$.
    \item $Q_{k}$ denotes the $k^{th}$ percentile. In case of $k = 90$, $Q_{90}$ is the $90^{th}$ percentile of the samples.    
    \item $\rho_{k}$ denotes the share of samples equal to or less than $k$.
    \item $\pi_{k}$ denotes the share of samples greater than $k$.
    \item $P$ denotes probability, e.g., $P(X \leq x) = k$ indicates that the probability of the random variable $X$ being equal to or less than $x$ is equal to $k$.
    \item $v(S)$ represents the worth of coalition $S$ ($S \in N$) in a cooperative game, and $\phi_i(v)$ denotes the Shapley value of the $i^{th}$ player.    
    \item $\alpha \in (0, 1)$ is a parameter used to adjust the metrics.
    \item $V_{\chi_i}, V_{\chi_o}, V_{\chi_f}$ denote the metric values for background traffic, the examined flow, and the overall traffic sum, respectively.
\end{itemize}

\subsection{Features}
\label{subsec:metric-features}

The characteristics of the proposed metrics are derived from the general concept introduced earlier. This research aims to propose a method for quantifying the impact of specific traffic components on network state over a given time period, with a particular focus on peak data rates. The proposed solution should either directly assess contributions to network load variations or enable impact assessment by comparing metric values across different time windows. For simplicity, the metric is assumed to be expressed as a single numerical value.

Given that metric properties can vary across use cases, providing a definitive set of requirements is challenging. However, in this study, metric selection is constrained by the runtime environment, including challenges in data collection and computational resource management. The requirement for metrics to be easily interpretable also dictates certain assumptions about the returned values. Based on these considerations, the following desired features have been identified:

\begin{enumerate}
    \item \textbf{passive data source}: the metric should use data collected through passive measurement techniques,
        \begin{itemize}
            \item using the metric must not involve generating traffic for active measurements,
            \item the mechanisms required for data collection are likely already implemented and operational in the network,
        \end{itemize}
    \item \textbf{easily collected input data}: the metric should require minimal input data,
        \begin{itemize}
            \item storing and processing the input data must not require excessive storage or computational resources,
        \end{itemize}
    \item \textbf{efficient computation}: the metric should be computable as quickly as possible,
        \begin{itemize}
            \item the metric should impose low computational resource requirements,
            \item computing metric values must not overload the control plane,
        \end{itemize}
    \item \textbf{single numerical output}: the metric should return a single floating-point value,
        \begin{itemize}
            \item the metric value should be easy to process,
        \end{itemize}
    \item \textbf{normalized within the range $[-1, 1]$}: the metric value should be normalized within the range of $-1$ to $1$,
        \begin{itemize}
            \item the metric should be easy to interpret and compare,
        \end{itemize}
    \item \textbf{lower metric values always indicate better network outcomes}: a value of $-1$ represents maximum improvement in network state, $0$ indicates no impact on network state, and $1$ denotes maximum network state degradation,
        \begin{itemize}
            \item comparing metric values should clearly indicate more desirable network effects.
        \end{itemize}
\end{enumerate}

\subsection{Base metrics}
\label{subsec:base-metrics}

Given its widespread use in passive measurement-based solutions, throughput is the sole \emph{base metric} considered in this paper. However, this metric undergoes further analysis in various ways to ensure diversity in the formulation of higher-tier metrics. The time series of \emph{base metrics} serves as the foundation for computing more complex \emph{derived metrics}. Future research may also explore extending the set to include additional metrics, such as latency and jitter~\cite{ponce:ecuador-internet-analysis}.

The throughput metric quantifies the amount of data transferred per unit of time, expressed in \ac{bps}. Metric values are typically collected periodically for each monitored link. For the measurements described in this paper, collecting throughput data in a network consisting of $|E|$ links over a $t$ period of time results in a data matrix $B$ of size $|E| \times w$, where $i$ represents the time interval and the number of sampling timestamps is given by $w = t/i$. The~value of $B_{l,w}$ represents the traffic throughput of the $l^{th}$ link at the $w^{th}$ timestamp. Example throughput values for two different network links are presented in Fig.~\ref{fig:example-throughput-time-series}. This metric serves as a valuable foundation for further network analysis, as it directly reflects activity on specific links.

\subsection{Derived metrics}
\label{subsec:derived-metrics}

The set of derived metrics includes link utilization, mean link utilization, and \ac{GFU}. These metrics are calculated using \emph{base metrics} along with additional data, such as link capacity and the active flow database, as input. The resulting data series represent network-wide measurements or per-link statistics within the topology.

\subsubsection{Link utilization}

The link utilization metric quantifies the proportion of link capacity in use, offering a simple method to distinguish between underutilized and overutilized links. While useful for basic monitoring against predefined thresholds, interpreting data from multiple links over extended periods remains challenging. Given the traffic throughput matrix $B$ of size ($|E| \times w$) and the link capacity vector $C$, the link utilization matrix is defined as $U = B/C$. The value of $U_{l,w}$ represents the link utilization of the $l^{th}$ link at timestamp $w$. 

\subsubsection{Mean link utilization}

The mean link utilization metric calculates average utilization values across all network links at each timestamp. While disregarding individual link data may obscure certain network events, this metric captures the predominant network state. It is stored in a vector $\overline{U}$ of length $w$. The value of $\overline{U}_{w} = \sum_l U_{l, w}/|E|$ represents the average utilization of all network links at timestamp $w$.

\subsubsection{Growth \acs{GFU}}

The Growth \ac{GFU} metric quantifies the network's ability to accommodate traffic and future flow growth. Its theoretical foundations and operational principles are detailed in the original research paper~\cite{hassidim:gfu}. The conclusions drawn suggest it as a promising alternative to mean link utilization for capacity planning. Therefore, this flow-oriented metric is considered a potential component of the solutions proposed in this paper. It should be noted that calculating metric values requires additional information on active flows, in addition to link utilization. Furthermore, computing the $\alpha^{growth}$ vector required to determine Growth \ac{GFU} can be time-consuming for large input datasets. Metric values are stored in a vector ${G}$ of length $w$. The Growth \ac{GFU} value at timestamp $w$ is denoted as ${G}_{w}$.

\subsubsection{Risk \acs{GFU}}

Risk \ac{GFU} is an approach similar to Growth \ac{GFU} that incorporates factors associated with flow failure risks~\cite{hassidim:gfu}. As a result, this metric offers valuable insights into network state. Analogously, the metric values are stored in a vector ${G}$ of length $w$. The Risk \ac{GFU} value at timestamp $w$ is denoted as ${G}_{w}$.

\subsection{Time series analysis}
\label{subsec:ts-analysis}

The time series of \emph{derived metrics} consist of a set of values associated with their respective timestamps. In subsequent steps, it is required that the characteristics of these datasets are expressed in a comprehensive way. One approach involves calculating the minimum, maximum, or average values of the samples. However, these methods are insufficient for analyzing traffic characteristics, as they discard important information about value distribution. Therefore, two alternative approaches are explored. The following analysis methods, categorized under \emph{network load metrics}, compute a single numerical value representing the entire input dataset. 

\subsubsection{Percentiles}

The $k^{th}$ percentile ($Q_k$) represents the value for which $P(X \leq Q_k) = k/100$. In network traffic measurements, analyzing the percentile value helps determine the maximum observed metric value while excluding a chosen percentage of the top samples. This approach effectively calculates baseline values that disregard minor traffic bursts for billing purposes.

\subsubsection{Top/bottom samples share}

The second analysis method quantifies the percentage of samples that fall below or exceed a specified threshold. Namely, $\rho_k$ represents the percentage share of samples that are equal to or smaller than $k$ (i.e., $P(X \leq k) = \rho_k$), referred to as bottom samples. Conversely, $\pi_k$ represents the percentage of samples greater than $k$ (i.e., $P(X > k) = \pi_k$), referred to as top samples. In network traffic measurements, these values can indicate the prevalence of exceptionally low or high metric readings and, for example, help classify the network state as underutilized or overutilized.

\subsection{Flow impact analysis}
\label{subsec:contrib-analysis}

\begin{figure}[t]
  \begin{center}
    \begin{tikzpicture}
      \begin{groupplot}[        
        table/col sep=comma,
        group style={
            group size=1 by 3,
            xlabels at=edge bottom,
            ylabels at=edge left,
            vertical sep=1.5cm
        },
        width=\columnwidth,
        height=0.5\columnwidth,
        xlabel={Time},        
        ylabel={Link utilization [1/1]},
        grid=major,
        ymin=0, ymax=1,
        xmajorticks=false,
        xminorticks=false
      ]

        \nextgroupplot[title={a) Initial background traffic.}]
        \addplot+[mark size=0.5pt] table[x=index,y=initial] {txs_example.csv};

        \nextgroupplot[title={b) Overlay flow traffic.}]
        \addplot+[mark size=0.5pt] table[x=index,y=overlay] {txs_example.csv};

        \nextgroupplot[title={c) Final overall traffic.}]
        \addplot+[mark size=0.5pt] table[x=index,y=final] {txs_example.csv};
      
      \end{groupplot}
    \end{tikzpicture}
    \caption{Example link utilization for initial background traffic, overlay flow and final overall traffic}
    \label{fig:example-txs}
  \end{center}
\end{figure}
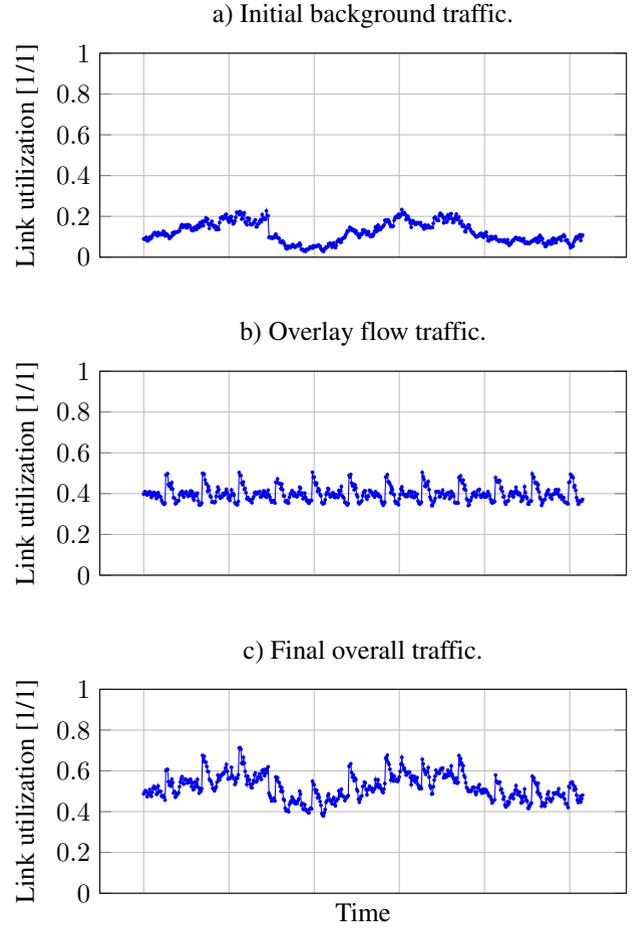

Network flows distinctly influence \emph{network load metrics}, which can be assessed through different metric values. Since overall network traffic (Fig.~\ref{fig:example-txs}c) comprises both the examined flow (Fig.~\ref{fig:example-txs}b) and background traffic (Fig.~\ref{fig:example-txs}a), metric values should be calculated in two different scenarios:

\begin{itemize}
    \item $V_{\chi_i}$ --- metric value for the network state excluding the examined flow from overall traffic,
    \item $V_{\chi_f}$ --- metric value for the network state including the examined flow in overall traffic.
\end{itemize}

Additionally, some metrics can be evaluated exclusively for the isolated traffic of the examined flow ($V_{\chi_o}$). Based on these three values, the impact of the flow on metric values can be determined using the following methods.

\subsubsection{Value difference (delta)}

The simplest method to assess the impact of a flow on a metric is to calculate the difference between its two values (Equation~\ref{eq:value-delta}). However, this approach may not always accurately reflect the flow's contribution to changes in the metric.

\begin{equation}
\label{eq:value-delta}
    \Delta = V_{\chi_f} - V_{\chi_i}
\end{equation}

\subsubsection{Shapley value}

The Shapley value ensures fair payoff distribution in cooperative games. With $n$ denoting the total players, $S$ representing a player coalition, and $v$ assigning expected payoff to a given coalition, the expected fair payoff for player $i$ (calculated as the average marginal contribution across all possible coalitions) is given by:

\begin{equation}
\label{eq:shapley}
\phi_i(v)=\sum_{S \subseteq N \setminus
\{i\}} \frac{|S|!\; (n-|S|-1)!}{n!}(v(S\cup\{i\})-v(S))    
\end{equation}

In network measurements, this concept can be applied to quantify the contribution of a single flow to a metric value. In this case, the total payoff corresponds to one of the previously described metrics, such as the $90^{th}$ percentile of link utilization samples calculated for overall traffic ($V_{\chi_f}$). To properly calculate the Shapley value, the metric value must also be evaluated for all possible coalitions, including the flow itself ($V_{\chi_o}$) and the initial network traffic alone ($V_{\chi_i}$), and applied to Equation~\ref{eq:shapley}. For multiple traffic components, applying the algorithm may be impractical due to its computational complexity, requiring approximation to maintain acceptable performance.

\begin{figure}[h!]
\centering
\begin{tikzpicture}[node distance=2.2cm, auto, font=\small]

\tikzstyle{startstop} = [rectangle, rounded corners, minimum width=3.5cm, minimum height=1.2cm, text centered, draw=black, fill=cyan!20]
\tikzstyle{process} = [rectangle, minimum width=4cm, minimum height=1cm, text centered, draw=black, fill=orange!30]
\tikzstyle{arrow} = [thick,->,>=stealth]

\node (start) [startstop] {Input: List of flows and network metric function $v(S)$};
\node (subset) [process, below of=start] {Generate all subsets $S \subseteq F$};
\node (marginal) [process, below of=subset] {Calculate marginal contributions: $v(S \cup \{f_i\}) - v(S)$};
\node (weights) [process, below of=marginal] {Assign weights to subsets based on size};
\node (aggregation) [process, below of=weights] {Aggregate contributions to compute $\phi_i$};
\node (output) [startstop, below of=aggregation, minimum height=2cm] {\parbox{3.5cm}{\centering Output: Shapley values $\phi_i$ for each flow. \\ List of flows with high $\phi_i$}};

\draw [arrow] (start) -- (subset);
\draw [arrow] (subset) -- (marginal);
\draw [arrow] (marginal) -- (weights);
\draw [arrow] (weights) -- (aggregation);
\draw [arrow] (aggregation) -- (output);

\end{tikzpicture}
\caption{Flowchart for Shapley value application in network traffic analysis}
\label{fig:shapley-flowchart}
\end{figure}
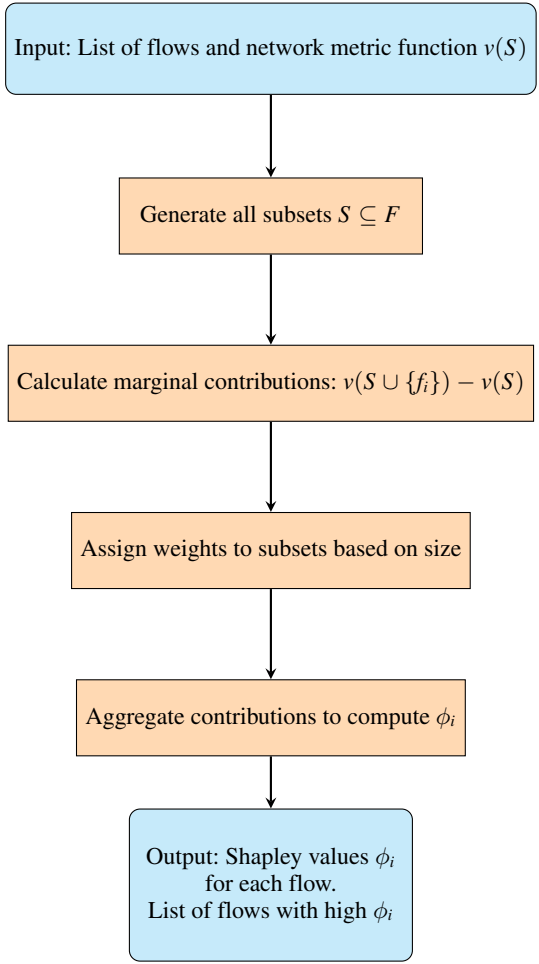

An example application of the Shapley value for identifying flows that impact network load is as follows. The implementation begins by defining the flow set $F = \{f_1, f_2, \dots, f_n\}$, and the coalition function $v(S)$, which represents the metric value for any subset $S \subseteq F$. All flow subsets are enumerated, and for each flow $f_i$, its marginal contribution to the network load within subset $S$ is calculated as $v(S \cup \{f_i\}) - v(S)$. Each contribution is weighted according to the coalition size, using the formula $|S|!(n - |S| - 1)!/n!$. Finally, the Shapley value for $f_i$ is computed by aggregating these weighted contributions across all subsets, as presented in Equation~\ref{eq:shapley}. The calculated Shapley values $\phi_i$ quantify each flow's contribution to the network's metric value. Higher values indicate a significant impact on the network and can guide flow selection for rerouting or rate limiting. A flowchart illustrating this example is presented in Fig.~\ref{fig:shapley-flowchart}.

\subsection{Summary}

This section has introduced the metrics and analysis methods used to construct \emph{network load metrics} and \emph{flow impact metrics}. A summary of these concepts is provided in Table~\ref{tab:metrics-methods-summary}, which includes both metric types, concept names, and references to related work.

\begin{table}
\centering
\caption{Metrics and analysis methods}
\label{tab:metrics-methods-summary}
\resizebox{\columnwidth}{!}{ 
\begin{tabular}{|l|l|l|}
\hline
\thead{Name}          & \thead{Type}                                                    & \thead{Related work} \\ \hline
Link utilization      & \multirow{4}{*}{Metric}                                         & \cite{hassidim:gfu} \\ \cline{1-1}\cline{3-3}
Mean link utilization &                                                                 & \cite{hassidim:gfu} \\ \cline{1-1}\cline{3-3}
Growth GFU            &                                                                 & \cite{hassidim:gfu} \\ \cline{1-1}\cline{3-3}
Risk GFU              &                                                                 & \cite{hassidim:gfu} \\ \hline
Percentile            & \multirow{2}{*}{\makecell[l]{Time series \\ analysis method}}   & \cite{vamseedhar:percentile, dimitropoulos:percentile, stanojevic:shapley, rzym:dtm} \\ \cline{1-1}\cline{3-3}
Top samples share     &                                                                 & - \\ \hline
Value difference      & \multirow{2}{*}{\makecell[l]{Impact \\ analysis method}}        &  \cite{zhang:evaluation} \\ \cline{1-1}\cline{3-3}
Shapley value         &                                                                 & \cite{stanojevic:shapley, vamseedhar:percentile} \\ \hline
\end{tabular}
}
\end{table}

\section{Network load and flow impact metrics}
\label{sec:load-impact-metrics}

This section presents the developed metrics that constitute the main contribution of this paper. The following subsections detail both the \emph{network load metrics} and \emph{flow impact metrics}, organized by their operational principles. The section concludes with a summary that establishes clear connections between the proposed \emph{flow impact metrics}, the existing literature, fundamental concepts, and the expected features outlined in Sections~\ref{sec:background} and \ref{sec:basic-metrics}.   

\subsection{Network load metrics}
\label{subsec:network-load-metrics}

The \emph{network load metrics} integrate \emph{derived metrics} with various analytical methods to represent an entire dataset of measurements as a single numerical value for a specific time window. These metrics, as presented in this paper, evaluate dataset percentiles or the distribution of extreme samples (top or bottom), while incorporating metrics such as link utilization and \ac{GFU}. Based on the analytical approaches employed, we organize the metrics into three distinct groups.

\subsubsection{Percentile-based metrics}

The first category of \emph{network load metrics} employs the percentile analysis method. Unlike related work that focuses on billing mechanisms using traffic volume measurements, these metrics leverage different \emph{derived metrics}: link utilization, mean link utilization, and \ac{GFU}.

Each metric calculates a single percentile value from its input time series, ensuring computational efficiency. The $\alpha$ parameter adjusts the calculated $[100\times(1 - \alpha)]^{th}$ percentile. For example, in the case of $\alpha = 0.10$, the value of $Q_{90}$ is calculated. Given the selected input metrics, the final metric value falls within the $[0, 1]$ range.

These metrics analyze the values of link utilization, mean link utilization, or \ac{GFU} within a defined measurement window. The result yields the highest observed metric value while systematically excluding the top traffic peaks according to the specified threshold. Consequently, the metric effectively determines the maximum traffic level after filtering out statistical outliers from the upper portion of the distribution. Lower metric values correspond to reduced network load observed during the measurement period, with the sensitivity to peak traffic controlled precisely by the $\alpha$ parameter.

The complete list of proposed \emph{network load metrics} based on percentile analysis is as follows:
\begin{itemize}
    \item Link Utilization Percentile (Listing~\ref{lst:lup})
    \item Mean Link Utilization Percentile (Listing~\ref{lst:mlup})
    \item Growth \ac{GFU} Percentile (Listing~\ref{lst:gfup})
    \item Risk \ac{GFU} Percentile (Listing~\ref{lst:gfup})
\end{itemize}

\begin{lstlisting}[caption={Link Utilization Percentile},label={lst:lup}]
Input: M[l][w] (link utilization)
       alpha (percentile adjustment)
Output: X (Link Utilization Percentile)

1. Flatten M into T and sort ascending

2. Calculate the percentile threshold:
    p = 1 - alpha
    i = ceil(p * len(T))
       
3. Calculate X:
    X = T[i]
\end{lstlisting}

\begin{lstlisting}[caption={Mean Link Utilization Percentile},label={lst:mlup}]
Input: M[l][w] (link utilization)
       alpha (percentile adjustment)
Output: X (Mean Link Utilization Percentile)

1. Calculate mean link util. A[w]:
    For each time interval w:       
        A[w] = mean(M[l][w] for all l)

2. Sort A ascending

3. Calculate the percentile threshold:
    p = 1 - alpha
    i = ceil(p * len(A))
       
4. Calculate X:
       X = A[i]
\end{lstlisting}

\begin{lstlisting}[caption={Growth \acs{GFU}/Risk \acs{GFU} Percentile},label={lst:gfup}]
Input: G[w] (GFU)
       alpha (percentile adjustment)
Output: X (GFU Percentile)

1. Sort G in ascending order

2. Calculate the percentile threshold:
    p = 1 - alpha
    i = ceil(p * len(G))
       
3. Calculate X:
    X = G[i]
\end{lstlisting}

\subsubsection{Top samples share-based metrics}

The second group of \emph{network load metrics} employs top samples share analysis, determining the percentage of samples exceeding specified thresholds. This approach analyzes the same set of derived metrics as the previous category: link utilization, mean link utilization, and \ac{GFU}.

Calculating the metric value for each input series is straightforward, primarily involving dataset sorting and identification of sample indices nearest to the threshold value. The threshold value in the metric is determined by the value of $(1 - \alpha)$. For example, when $\alpha = 0.10$, the algorithm calculates the value of $\pi_{90}$. The resulting metric value always falls within the range $[0, 1]$.

These metrics identify instances where defined thresholds are exceeded within a specified time window. This information enables network administrators to determine whether utilization levels remain within nominal ranges and quantify the frequency of threshold violations. Consequently, these metrics serve as effective tools for detecting the overutilization of network resources. Lower metric values indicate less frequent periods of overutilization, with a value of 0 representing the optimal case where no overutilization is observed.

The complete list of developed \emph{network load metrics} based on top samples analysis is as follows:
\begin{itemize}
    \item Top Link Utilization Samples Share (Listing~\ref{lst:tluss})
    \item Top Mean Link Utilization Samples Share (Listing~\ref{lst:tmluss})
    \item Top Growth \ac{GFU} Samples Share (Listing~\ref{lst:tgss})
    \item Top Risk \ac{GFU} Samples Share (Listing~\ref{lst:tgss})
\end{itemize}

\begin{lstlisting}[mathescape=true, caption=Top Link Utilization Samples Share,label={lst:tluss}]
Input: M[l][w] (link utilization)
       alpha (threshold adjustment)
Output: C (Top Link Utilization Samples Share)

1. Flatten M into T

2. Calculate the threshold:
    th = 1 - alpha
       
3. Count v. exceeding the threshold:
    C = count(v > th for v in T)
           
4. Calculate X:
    X = C / len(T)
\end{lstlisting}

\begin{lstlisting}[mathescape=true, caption=Top Mean Link Utilization Samples Share,label={lst:tmluss}]
Input: M[l][w] (link utilization)
       alpha (threshold adjustment)
Output: X (Top Mean Link Utilization Samples Share)

1. Calculate mean link util. A[w]:
    For each time interval w:       
        A[w] = mean(M[l][w] for all l)

2. Calculate the threshold:
    th = 1 - alpha
       
3. Count v. exceeding the threshold:
    C = count(v > th for v in A)
           
4. Calculate X:
    X = C / len(A)
\end{lstlisting}

\begin{lstlisting}[caption={Top Growth \acs{GFU}/Risk \acs{GFU} Samples Share},label={lst:tgss}]
Input: G[w] (GFU)
       alpha (threshold adjustment)
Output: X (Top GFU Samples Share)

1. Calculate the threshold:
    th = 1 - alpha
       
2. Count v. exceeding the threshold:
    C = count(g > th for g in G)
           
3. Calculate X:
    X = C / len(G)
\end{lstlisting}

\subsubsection{Utilization Score}

The Utilization Score, designed specifically for this study, simultaneously evaluates both overutilization and underutilization of network resources. The threshold utilization levels are defined relative to the $\alpha$ parameter: metric values less than or equal to $\alpha$ indicate underutilization, while values greater than $1 - \alpha$ represent overutilization. The metric incorporates both sample types into a weighted average, assigning a higher weight to overutilized samples. This design decision ensures the metric remains more sensitive to overutilization, as exceeding critical load thresholds can significantly disrupt network traffic. Conversely, underutilization consequences are considerably less severe, as excess provisioned resources do not threaten network performance or stability. The formula for the Utilization Score is presented in Equation~\ref{eq:uscore}, with the corresponding implementation pseudocode provided in Listing~\ref{lst:uscore}.

\begin{equation}
\label{eq:uscore}
\mathit{US} = \frac{1000 \times \pi_{1 - \alpha} + 100 \times \rho_{\alpha}}{1000}
\end{equation}

The Utilization Score produces values within the range $[0, 1]$, where lower values indicate more desirable network conditions. An optimal Utilization Score of 0 indicates that throughout the entire measurement period, all metric values remained within the nominal range $(\alpha, 1-\alpha]$.

\begin{lstlisting}[caption=Utilization Score, label={lst:uscore}]
Input: M[l][w] (link utilization)
       alpha (threshold adjustment)
Output: X (Utilization Score)

1. Calculate the thresholds:
    t_u = alpha  // underutil.
    t_o = 1 - alpha  // overutil.

2. Count under- and overutilized val.:
    C_u = sum(M[l][w]<=t_u for all l,w)
    C_o = sum(M[l][w]>t_o for all l,w)

3. Calculate the shares:
    s_u = C_u / len(M)
    s_o = C_o / len(M)

4. Calculate X:
    X = (1000 * s_o + 100 * s_u) / 1000
\end{lstlisting}

\subsection{Flow impact metrics}
\label{subsec:flow-impact-metrics}

The \emph{flow impact metrics} constitute the final contribution of this paper and quantify how individual flows affect the values of \emph{network load metrics}. To accomplish this objective, we apply \emph{contribution analysis methods} to \emph{network load metric} values calculated from paired datasets --- one including and one excluding the examined flow. The proposed metrics are organized into two groups based on their underlying analysis methods.

\subsubsection{Value difference (delta)-based metrics}

The first group of metrics calculates differences between \emph{network load metric} values across comparative datasets. Despite its simplicity, this method effectively identifies changes in metric values with minimal computational overhead. Listing~\ref{lst:delta} presents the pseudocode implementation of this calculation method. Since the underlying \emph{network load metrics} are normalized within the range $[0, 1]$, the resulting difference falls between $-1$ (representing the highest positive impact) and $1$ (indicating the highest negative impact).

The complete list of \emph{flow impact metrics} based on the value difference method is as follows:
\begin{itemize}
    \item \ac{LUPD}
    \item \ac{MLUPD}
    \item \ac{GGPD}
    \item \ac{TLUSSD}
    \item \ac{TMLUSSD}
    \item \ac{TGGSSD}
    \item \ac{LUSD}
    \item \ac{RGPD}
    \item \ac{TRGSSD}
\end{itemize}

\begin{figure}[!t]
\centering
\begin{lstlisting}[caption=Value Difference (Delta)-based Metric Calculation,label={lst:delta}]
Input: M[l][w] (link utilization)
       alpha (threshold adjustment)
Output: D (flow impact)

1. Calculate the metric values:
    V_xi $\leftarrow$ background traffic only
    V_xf $\leftarrow$ overall traffic

2. Calculate the v. difference (delta):
    D = V_xf - V_xi
\end{lstlisting}
\end{figure}

\subsubsection{Normalized Shapley share-based metrics}

The second group consists of metrics based on the Shapley value analysis method. While previous research primarily applied the Shapley value to traffic volume percentiles, this paper extends the approach to two distinct properties of link utilization samples: percentiles and top sample shares.

These metrics enable direct assessment of how specific traffic components (particularly selected flows) contribute to \emph{network load metric} values. The approach defines three coalitions: the examined flow in isolation, the background traffic without the examined flow, and the combined overall traffic.

Our approach modifies the standard Shapley method through a two-step normalization process. In the first step, we calculate the ratio between the computed Shapley value and the total payoff (denoted by $v(N)$, where $N$ represents the coalition of all players) as shown in Equation~\ref{eq:shapley-normalization-1}. In the second step, we apply a significance threshold, recognizing that only values representing major contributions (exceeding 50\%) to the metric value should be considered. Consequently, the metric is restricted to values in the range $[0, 0.5]$ using the Heaviside step function (Equation~\ref{eq:heaviside-function}) and subsequently rescaled to the standard $[0, 1]$ range (Equation~\ref{eq:shapley-normalization-2}). As with the value difference-based metrics, lower values in this metric class indicate reduced impact on the overall network state.

\begin{equation}
\label{eq:shapley-normalization-1}
\displaystyle \phi' = {\begin{cases}\frac{\phi}{v(N)},&v(N)>0\\0,&v(N)\leq 0\end{cases}}
\end{equation}

\begin{equation}
\label{eq:heaviside-function}
\displaystyle H(x) \equiv {\begin{cases}1,&x\geq 0\\0,&x<0\end{cases}}
\end{equation}

\begin{equation}
\label{eq:shapley-normalization-2}
\phi'' = 2 \times (\phi' - 0.5) \times H(\phi' - 0.5)
\end{equation}

While existing research suggests that Shapley value implementation in billing mechanisms can be computationally intensive, our approach significantly reduces this complexity by considering only two players and a fixed, minimal number of possible coalitions. Consequently, this approach enables efficient metric value computation with minimal computational overhead. The implementation of our Shapley value calculation method is presented in Listing~\ref{lst:shapley}.

The requirement of computing metric values specifically for isolated flow traffic ($V_{\chi_o}$) imposes certain limitations on the applicable \emph{network load metrics}. Consequently, the complete list of \emph{flow impact metrics} derived using the Shapley value method is as follows:
\begin{itemize}
    \item \ac{LUPSV} (normalized)
    \item \ac{TLUSSSV} (normalized)
\end{itemize}

\begin{figure}[!t]
\centering
\begin{lstlisting}[mathescape=true, caption=Normalized Shapley Share-based Metric Calculation,label={lst:shapley}]
Input: M[l][w] (link utilization)
       v(N) (total payoff for the coalition of all players)
       alpha (threshold adjustment)
Output: F (flow impact)

1. Calculate the metric values:
    V_xi $\leftarrow$ background traffic only
    V_xo $\leftarrow$ examined flow only
    V_xf $\leftarrow$ overall traffic

2. Calculate the Shapley value:
    F_f = (V_xf - V_xi + V_xo) / 2

3. Normalize the metric value:
    F_n = F_f / v(N) if v(N) > 0 else 0
    F = 2*(F_n-0.5) if F_n > 0.5 else 0
\end{lstlisting}
\end{figure}

\subsection{Summary}

This section has introduced nine \emph{network load metrics}, based on diverse measurement approaches. These metrics employ various analytical techniques, including percentile values and top samples share methodologies, applied to both link utilization and \ac{GFU}. Among these, the Utilization Score metric represents our original contribution to the research. We developed 11 \emph{flow impact metrics} to quantify how individual flows affect network state through their influence on metric values. Table~\ref{tab:flow-impact-metrics-summary} provides a comprehensive summary of these metrics and their corresponding analysis methods.

\begin{table*}
\caption{Flow impact metrics}
\label{tab:flow-impact-metrics-summary}
\centering
\resizebox{\textwidth}{!}{ 
\renewcommand{\arraystretch}{1.2} 
\begin{tabular}{|l|u|u|u|u|v|v|v|c|c|}

\hline
 \multirow{2}{*}{\thead{Name}} & 
 \multicolumn{4}{c|}{\thead{Derived metric}} &
 \multicolumn{3}{c|}{\thead{Network load metric}} &
 \multicolumn{2}{c|}{\thead{Flow impact \\ analysis}}
\\ \cline{2-10}
 &
    \thead{Link \\ utilization} &
    \thead{Mean \\ link utilization} &
    \thead{Growth \\ \acs{GFU}} &
    \thead{Risk \\ \acs{GFU}} &
    \thead{Percentile} &
    \thead{Top/bottom \\ samples share} &    
    \thead{Utilization \\ Score} &
    \thead{Delta} &
    \thead{Shapley \\ Value} \\ \hline
\ac{LUPD}     &    \markyes & & & & \markyes & & & \markyes &                   \\ \hline
\ac{TLUSSD}   &     \markyes & & & &  & \markyes & & \markyes &                 \\ \hline
\ac{LUSD}     &     \markyes & & & & & \markyes & \markyes & \markyes &         \\ \hline
\ac{LUPSV}    &     \markyes & & & &    \markyes & & &  & \markyes              \\ \hline
\ac{TLUSSSV}  &     \markyes & & & &  & \markyes & & & \markyes                 \\ \hline
\ac{MLUPD}    & &   \markyes & & &      \markyes & & & \markyes &               \\ \hline
\ac{TMLUSSD}  & &   \markyes & & & &    \markyes & & \markyes &                 \\ \hline
\ac{GGPD}     &  & & \markyes & & \markyes & & & \markyes &                     \\ \hline
\ac{TGGSSD}   &  & & \markyes & & & \markyes & & \markyes &                     \\ \hline
\ac{RGPD}     &  & &  & \markyes & \markyes & & & \markyes &                    \\ \hline
\ac{TRGSSD}   &  & &  & \markyes & & \markyes & & \markyes &                    \\ \hline
\end{tabular}
}
\end{table*}

\section{Evaluation}
\label{sec:experiments}

This section presents the experimental evaluation of the proposed \emph{flow impact metrics}. The evaluation begins with a brief outline of performance and functional characteristics. Next, we describe the input datasets and explain the composition of our test scenarios. Finally, we analyze the evaluation results with comprehensive commentary on metric behavior across various network conditions. The section concludes with a summary of key findings and their implications for metric selection and application.

\subsection{Performance and features}
\label{subsec:evaluation-performance}

We measured the computational performance for each metric and analysis method using input data extracted from one experimental dataset, which included:
\begin{itemize}
    \item \textbf{Link utilization}: $516 \times 36$-element matrix of throughput measurements and a 36-element vector of link capacities (network topology with 36 links and 516 measurement timestamps), 
    \item \textbf{Mean link utilization}: $516 \times 36$-element matrix of throughput measurements and a 36-element vector of link capacities,
    \item \textbf{Growth \ac{GFU}}: $516 \times 36$-element matrix of throughput measurements, a 53129-element list of flows and a 36-element vector of link capacities (ca. 100 flows per measurement timestamp),
    \item \textbf{Risk \ac{GFU}}: $516 \times 36$-element matrix of throughput measurements, a 53129-element list of flows and a 36-element vector of link capacities (ca. 100 flows per measurement timestamp),
    \item \textbf{Percentile analysis method}: $516 \times 36$-element matrix of link saturation,
    \item \textbf{Top samples share analysis method}: $516 \times 36$-element matrix of link saturation,
    \item \textbf{Value difference contribution analysis method}: two values of an example \emph{network load metric},
    \item \textbf{Shapley value contribution analysis method}: payoff values for 3 coalitions (a game with two players).
\end{itemize}

Table~\ref{tab:performance-results} presents the results rounded to three decimal places. Nearly all observed calculation times were negligible (close to $0$ seconds), with two notable exceptions: Risk \ac{GFU} (approx. 23\,s) and Growth \ac{GFU} (approx. 481\,s). The considerably longer computation time for Growth \ac{GFU} stems from the algorithmic complexity of calculating the admissible $\alpha^{growth}$ vector for each timestamp in the dataset.

\begin{table}
\centering
\caption{Performance results}
\label{tab:performance-results}
\resizebox{\columnwidth}{!}{ 
\renewcommand{\arraystretch}{1.2} 
\begin{tabular}{|l|l|S[table-format=3.3]|}
\hline\thead{Name}    & \thead{Type}                                                    & {\thead{Calculation \\ time [s]}}   \\ \hline
Link utilization      & \multirow{4}{*}{Metric}                                         & 0.001                               \\ \cline{1-1}\cline{3-3}
Mean link utilization &                                                                 & 0.001                               \\ \cline{1-1}\cline{3-3}
Growth GFU            &                                                                 & 481.461                             \\ \cline{1-1}\cline{3-3}
Risk GFU              &                                                                 & 22.973                              \\ \hline
Percentile            & \multirow{2}{*}{\makecell[l]{Time series \\analysis method}}      & 0                                   \\ \cline{1-1}\cline{3-3}
Top samples share     &                                                                 & 0                                   \\ \hline
Value difference      & \multirow{2}{*}{\makecell[l]{Impact \\analysis method}}           & 0                                   \\ \cline{1-1}\cline{3-3}
Shapley value         &                                                                 & 0.017                               \\ \hline
\end{tabular}
}
\end{table}

These performance considerations significantly influence the practical applicability of each metric. Table~\ref{tab:flow-impact-metrics-features} summarizes how the \emph{flow impact metrics} align with the desired characteristics outlined in Section~\ref{subsec:metric-features}. Except for the metrics based on \ac{GFU}, all metrics satisfy the established requirements. The four metrics --- \ac{GGPD}, \ac{TGGSSD}, \ac{RGPD}, and \ac{TRGSSD} --- exhibit long computation times that significantly limit their practical usability. Additionally, these metrics require comprehensive network flow information as input. This requirement substantially increases both storage demands and processing overhead for metric computation. It should be noted that the two metrics based on the Shapley value, \ac{LUPSV} and \ac{TLUSSSV}, have distinct characteristics. These metrics produce values within the range $[0, 1]$, where $0$ represents no detectable flow impact and $1$ represents the maximum possible impact on network load.

\begin{table}
\begin{center}    

\caption{Flow impact metrics features}
\label{tab:flow-impact-metrics-features}
\resizebox{\columnwidth}{!}{ 
\begin{tabular}{|l|c|c|c|c|c|c|}
\hline
\multirow{2}{*}{\thead{Name}} & \multicolumn{6}{c|}{\thead{Desired features}}        \\ \cline{2-7}
             & \thead{1}& \thead{2}& \thead{3}& \thead{4}& \thead{5} & \thead{6}     \\ \hline
\ac{LUPD}    & \markyes & \markyes & \markyes & \markyes & \markyes  & \markyes      \\ \hline
\ac{TLUSSD}  & \markyes & \markyes & \markyes & \markyes & \markyes  & \markyes      \\ \hline
\ac{LUSD}    & \markyes & \markyes & \markyes & \markyes & \markyes  & \markyes      \\ \hline
\ac{LUPSV}   & \markyes & \markyes & \markyes & \markyes & \markyes* & \markyes     \\ \hline
\ac{TLUSSSV} & \markyes & \markyes & \markyes & \markyes & \markyes* & \markyes     \\ \hline
\ac{MLUPD}   & \markyes & \markyes & \markyes & \markyes & \markyes  & \markyes      \\ \hline
\ac{TMLUSSD} & \markyes & \markyes & \markyes & \markyes & \markyes  & \markyes      \\ \hline
\ac{GGPD}    & \markyes &          &          & \markyes & \markyes  & \markyes      \\ \hline
\ac{TGGSSD}  & \markyes &          &          & \markyes & \markyes  & \markyes      \\ \hline
\ac{RGPD}    & \markyes &          &          & \markyes & \markyes  & \markyes      \\ \hline
\ac{TRGSSD}  & \markyes &          &          & \markyes & \markyes  & \markyes      \\ \hline
\end{tabular}
}
\end{center}

\vspace{4pt} 
\noindent
\textbf{1}: passive data source, \textbf{2}: easily collected input data, \textbf{3}: efficient computation, \textbf{4}: single numerical output, \textbf{5}: normalized within the range $[-1, 1]$, \textbf{6}: lower metric values always indicate better network outcomes

\textbf{*}: normalized within the range $[0, 1]$, additional notes in Section~\ref{subsec:evaluation-performance}

\end{table}

\subsection{Datasets}

We evaluated the metrics across three distinct network topologies: \emph{geant}, \emph{abilene}, and \emph{polska}, each featuring distinct background traffic patterns. The evaluation utilized pre-recorded traffic trace datasets derived from research studies detailed in subsequent sections. For each topology, we generated nine distinct test batches with varying flow characteristics, each stored in separate datasets.

\subsubsection{Topologies and background traffic}

\begin{figure}
    \centering
    \includegraphics[width=\columnwidth]{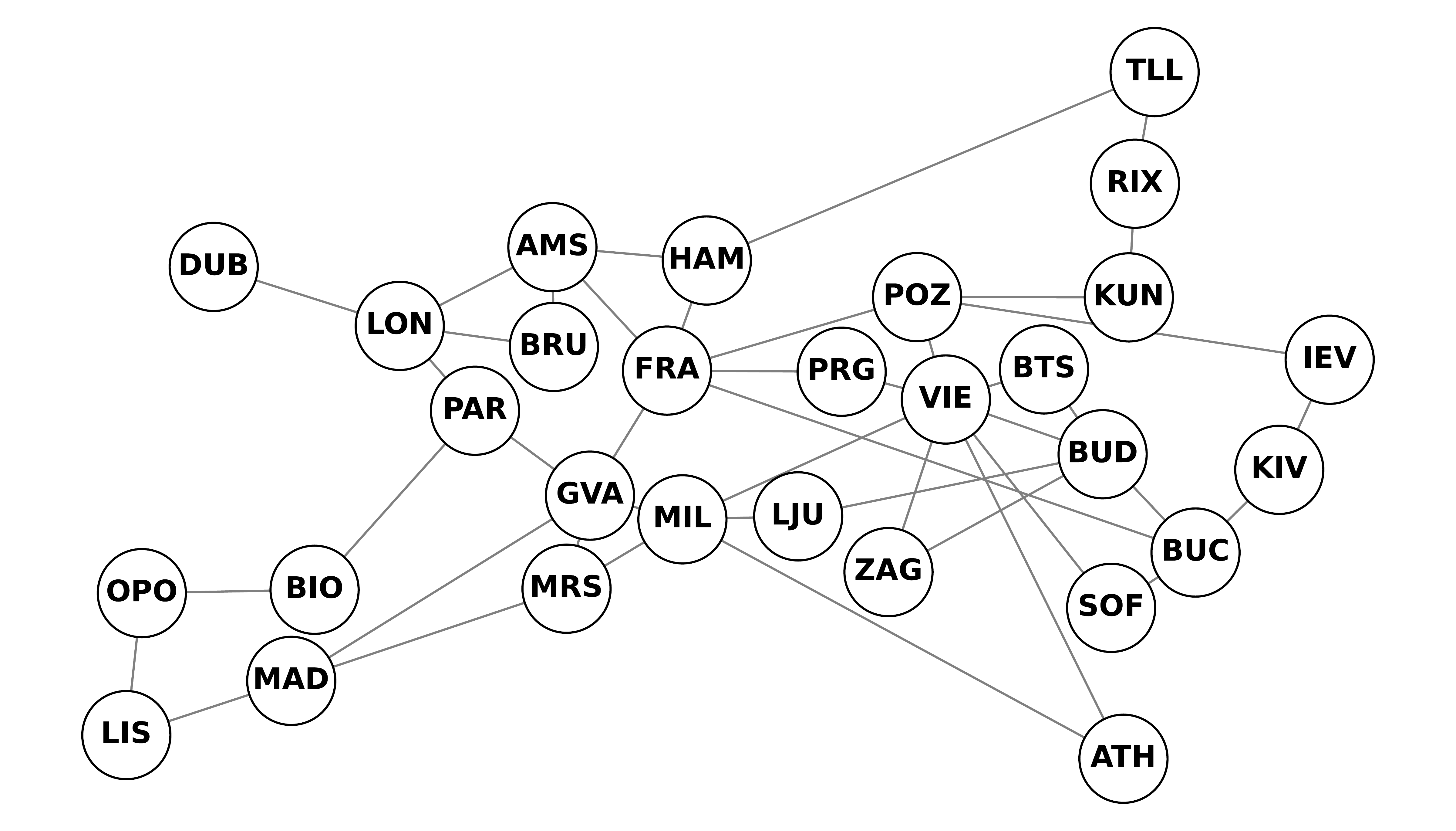}
    \caption{Topology of the \emph{geant} network ($|V| = 29, |E| = 88$)}
    \label{fig:geant-topo}
\end{figure}

The \emph{geant} topology replicates the European research network \ac{GÉANT}. The topology consists of 29 nodes and 88 links with capacities ranging from 25\,Gbps to 300\,Gbps. We retrieved traffic traces from the \ac{BRIAN} system~\cite{geant:brian} through automated \ac{API} queries sent to its Grafana application instance. Fig.~\ref{fig:geant-topo} illustrates this topology, while Table~\ref{tab:geant-capacity} details the individual link capacities.

\begin{table}[ht]
\begin{center}
\caption{Link capacities in the \emph{geant} topology}
\label{tab:geant-capacity}
\resizebox{\columnwidth}{!}{ 
\renewcommand{\arraystretch}{1.2} 
\begin{tabular}{|q|l|q|l|}
    \hline
    \rowcolor{white}
    \thead{Link}  & \thead{Capacity \\ \text{[Gbps]}} & \thead{Link}  & \thead{Capacity \\ \text{[Gbps]}} \\ 
    \hline
    AMS-BRU $\leftrightarrow$ & 25.00 Gbps   & MAD-MRS $\leftrightarrow$ & 100.00 Gbps \\ \hline   
    BTS-VIE $\leftrightarrow$ & 40.00 Gbps   & MIL-ATH $\leftrightarrow$ & 40.00 Gbps  \\ \hline
    BUC-FRA $\leftrightarrow$ & 40.00 Gbps   & MIL-GVA $\leftrightarrow$ & 200.00 Gbps \\ \hline
    BUD-BTS $\leftrightarrow$ & 40.00 Gbps   & MRS-GVA $\leftrightarrow$ & 200.00 Gbps \\ \hline
    BUD-BUC $\leftrightarrow$ & 40.00 Gbps   & MRS-MIL $\leftrightarrow$ & 100.00 Gbps \\ \hline
    BUD-VIE $\leftrightarrow$ & 200.00 Gbps  & OPO-BIO $\leftrightarrow$ & 200.00 Gbps \\ \hline
    DUB-LON                   & 200.00 Gbps  & PAR-BIO $\leftrightarrow$ & 200.00 Gbps \\ \hline
    LON-DUB                   & 100.00 Gbps  & PAR-GVA $\leftrightarrow$ & 300.00 Gbps \\ \hline
    FRA-AMS $\leftrightarrow$ & 200.00 Gbps  & PAR-LON $\leftrightarrow$ & 300.00 Gbps \\ \hline
    GVA-FRA $\leftrightarrow$ & 300.00 Gbps  & POZ-FRA $\leftrightarrow$ & 40.00 Gbps  \\ \hline
    HAM-AMS $\leftrightarrow$ & 200.00 Gbps  & PRG-FRA $\leftrightarrow$ & 200.00 Gbps \\ \hline
    HAM-FRA $\leftrightarrow$ & 200.00 Gbps  & RIX-KUN $\leftrightarrow$ & 40.00 Gbps  \\ \hline
    IEV-KIV $\leftrightarrow$ & 100.00 Gbps  & RIX-TLL $\leftrightarrow$ & 40.00 Gbps  \\ \hline
    IEV-POZ $\leftrightarrow$ & 100.00 Gbps  & SOF-BUC $\leftrightarrow$ & 40.00 Gbps  \\ \hline
    KIV-BUC $\leftrightarrow$ & 40.00 Gbps   & SOF-VIE $\leftrightarrow$ & 40.00 Gbps  \\ \hline
    KUN-POZ $\leftrightarrow$ & 40.00 Gbps   & TLL-HAM $\leftrightarrow$ & 40.00 Gbps  \\ \hline
    LIS-OPO $\leftrightarrow$ & 100.00 Gbps  & VIE-ATH $\leftrightarrow$ & 40.00 Gbps  \\ \hline
    LJU-BUD $\leftrightarrow$ & 100.00 Gbps  & VIE-MIL $\leftrightarrow$ & 100.00 Gbps \\ \hline
    LJU-MIL $\leftrightarrow$ & 200.00 Gbps  & VIE-POZ $\leftrightarrow$ & 100.00 Gbps \\ \hline
    LON-AMS $\leftrightarrow$ & 200.00 Gbps  & VIE-PRG $\leftrightarrow$ & 200.00 Gbps \\ \hline
    LON-BRU $\leftrightarrow$ & 25.00 Gbps   & ZAG-BUD $\leftrightarrow$ & 40.00 Gbps  \\ \hline    
    MAD-GVA $\leftrightarrow$ & 100.00 Gbps  & ZAG-VIE $\leftrightarrow$ & 40.00 Gbps  \\ \hline
    MAD-LIS $\leftrightarrow$ & 200.00 Gbps  & \multicolumn{2}{l}{}                    \\ \cline{1-2}

\end{tabular}
}
\end{center}

\vspace{4pt} 
\noindent
$\leftrightarrow$: same capacity for the link in the opposite direction 
\end{table}

\begin{figure}
    \centering
    \includegraphics[width=\columnwidth]{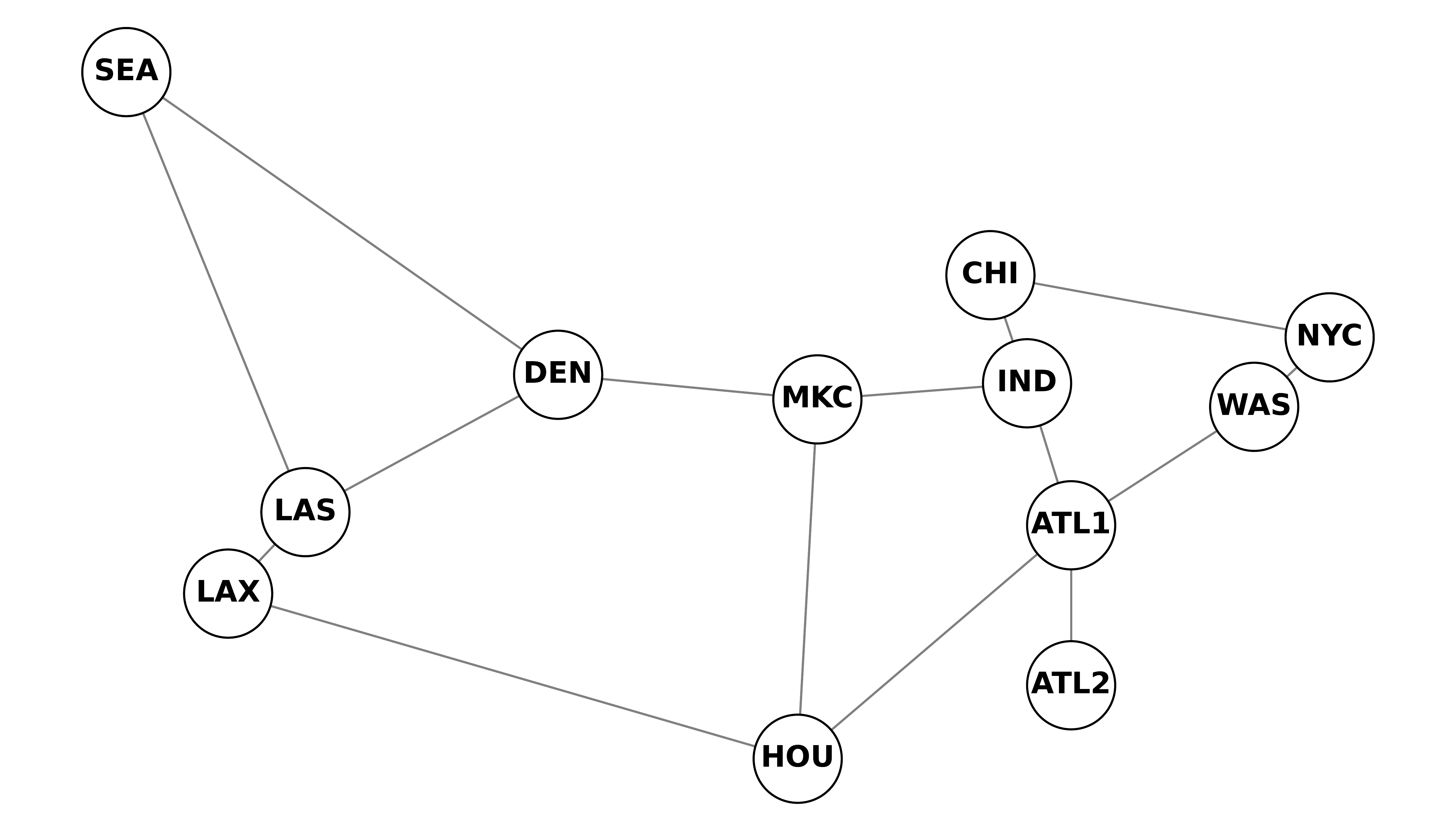}
    \caption{Topology of the \emph{abilene} network ($|V| = 12, |E| = 30$)}
    \label{fig:abilene-topo}
\end{figure}

The \emph{abilene} topology has been sourced from the SNDlib project~\cite{orlowski:sndlib}. The network consists of 12 nodes and 30 links with capacities between 7\,Gbps and 10\,Gbps. Specifically, links between nodes IND and ATL1 have 7.44\,Gbps capacity, while all other links operate at 9.44\,Gbps. Fig.~\ref{fig:abilene-topo} depicts the network topology. We derived traffic traces from demand matrices based on measurements from this US intercity backbone network. We applied shortest path routing to determine demand paths and consequently calculate link traffic patterns throughout the topology.

\begin{figure}
    \centering
    \includegraphics[width=\columnwidth]{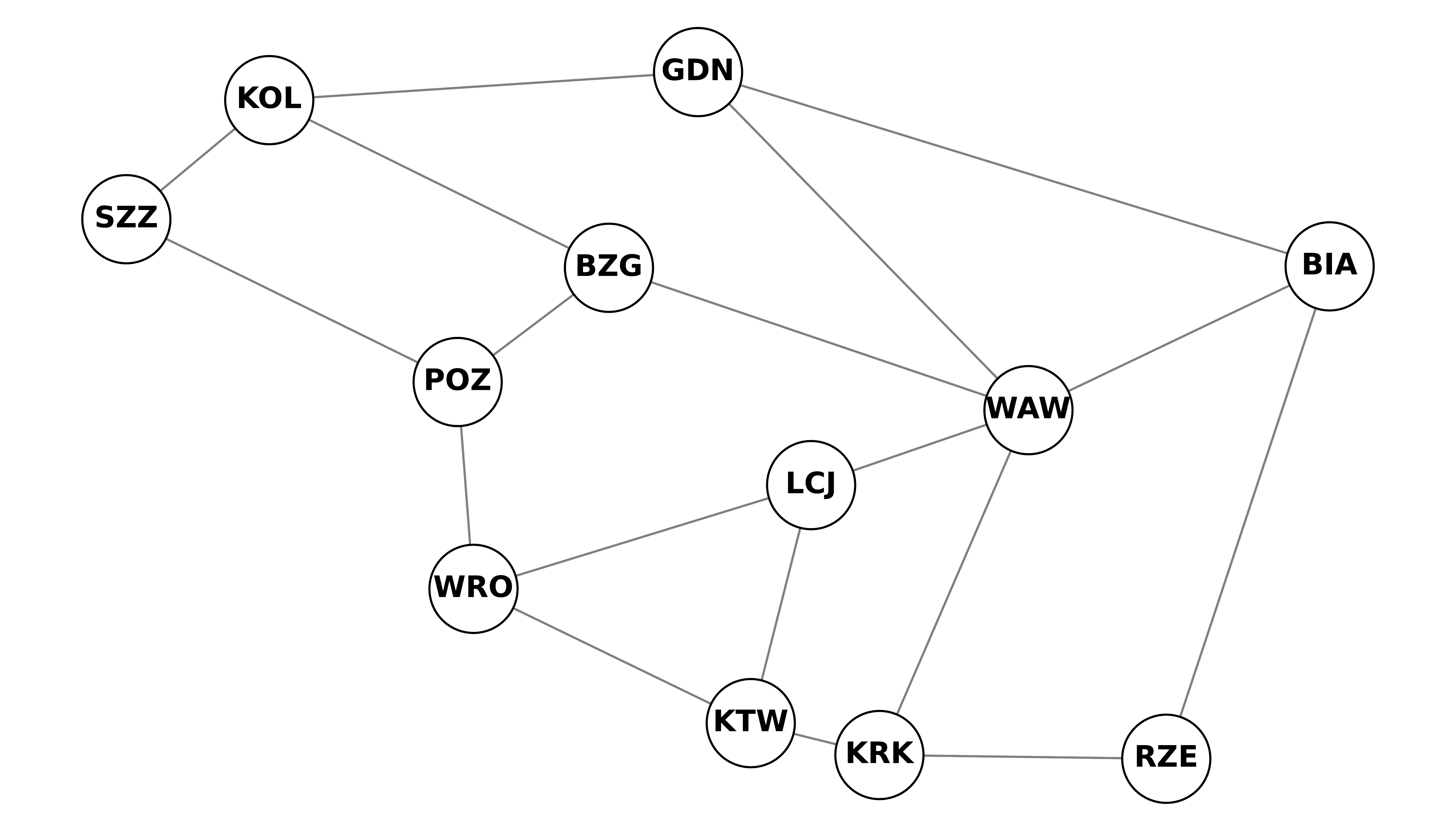}
    \caption{Topology of the \emph{polska} network ($|V| = 12, |E| = 36$)}
    \label{fig:polska-topo}
\end{figure}

The \emph{polska} topology derives from datasets developed during the SDNRoute project~\cite{borylo:sdnroute}. The network comprises 12 nodes and 36 links, as illustrated in Fig.~\ref{fig:polska-topo}. All links operate at a uniform capacity of 100\,Mbps. These datasets provide both structural information and traffic patterns. As part of the SDNRoute project, researchers developed a traffic model based on 30 days of measurements collected at AGH University of Krakow~\cite{borylo:sdnroute2}. This model generated realistic Internet-like traffic patterns for evaluating the SDNRoute system, and we utilized these resulting traffic traces in our current study.

From each topology, we extracted approximately two-hour traffic trace segments. To satisfy requirements for specific metrics, we either loaded pre-existing flow databases (for \emph{abilene}) or programmatically derived flow information by decomposing traffic patterns using custom scripts (for \emph{geant} and \emph{polska}).

\subsubsection{Test flows}

We extracted test flow data from YouTube mobile streaming traffic measurements documented in~\cite{karagkioules:yt}. The original dataset contains traffic measurements from YouTube's mobile streaming application under various conditions. Each trace was captured using the \emph{tcpdump} tool and stored in separate files labeled with scenario number, video ID, and iteration number. We developed a Python script to consolidate approximately 1,100 trace files into a unified data matrix, which served as the source for generating synthetic test flows across our experimental batches.

\subsubsection{Test batches}

We created a total of 27 experimental batches, allocating nine batches for each topology. Our batch generation process combines initial network background traffic with synthetically generated test flows. These test flows derive from the external dataset described in the previous section. For each batch, we adjusted the aggregated flows to match specific target parameters: mean bandwidth and standard deviation. As illustrated in Fig.~\ref{fig:example-txs}, we added the resulting test flow patterns (Fig.\ref{fig:example-txs}b) to the baseline traffic in each network link (Fig.~\ref{fig:example-txs}a), producing the composite traffic patterns for analysis (Fig.~\ref{fig:example-txs}c).

To evaluate metric performance across diverse network conditions, we defined distinct parameter sets for each batch. The primary parameter dimension was the mean bandwidth of the added test flows:
\begin{itemize}
    \item \emph{L} --- test flows with the lowest added test flow bandwidth,
    \item \emph{M} --- test flows with the medium added test flow bandwidth,
    \item \emph{H} --- test flows with the highest added test flow bandwidth.
\end{itemize}

The second parameter dimension was the standard deviation of test flow bandwidth:
\begin{itemize}
    \item \emph{A} --- test flows with the lowest standard deviation,
    \item \emph{B} --- test flows with the medium standard deviation,
    \item \emph{C} --- test flows with the highest standard deviation.
\end{itemize}

Importantly, despite having different standard deviations, flows in batches with identical mean bandwidth values transferred equal total data volumes. The additional batch label \emph{I} designates the initial state of the network prior to the introduction of any test flows. The notation system for referencing these datasets follows these conventions:
\begin{itemize}
    \item \emph{M batches} --- all batches with added medium standard deviation test flows,
    \item \emph{abilene-I} --- initial state of the traffic in the abilene topology,
    \item \emph{geant-M-A} --- the \emph{geant} topology with added test flows (medium bandwidth and lowest standard deviation).
\end{itemize}

Table~\ref{tab:evaluation-parameters} summarizes the complete parameter configuration for all datasets.

\begin{table}
\begin{center}
\caption{Summary of datasets}
\label{tab:evaluation-parameters}
\renewcommand{\arraystretch}{1.2} 
\begin{tabular}{|l|l|l|l|l|l|l|}
\hline
\thead{Topology} & \multicolumn{3}{c|}{\thead{Mean \\bandwidth [bps]}} & \multicolumn{3}{c|}{\thead{Standard deviation [bps]}} \\
\cline{2-7}
& \thead{L} & \thead{M} & \thead{H} & \thead{A} & \thead{B} & \thead{C} \\
\hline
polska & 24 M & 48 M & 72 M & 3 M & 6 M & 9 M \\ \hline
geant & 7 G & 14 G & 21 G & 1 G & 2 G & 3 G \\ \hline
abilene & 2 G & 4 G & 6 G & 300 M & 600 M & 900 M \\
\hline
\end{tabular}
\end{center}

\vspace{4pt} 
\noindent
\textbf{L}: low bandwidth, \textbf{M}: medium bandwidth, \textbf{H}: high bandwidth \\
\textbf{A}: low std. dev., \textbf{B}: medium std. dev., \textbf{C}: high std. dev.

\end{table}

\subsection{Implementation}

Our implementation process encompassed several key tasks to facilitate dataset preparation and metric evaluation:

\begin{itemize}
    \item retrieving traffic traces and topology from Grafana \ac{API} in \ac{BRIAN} system,
    \item loading traffic traces and topology from files provided by SNDlib and SDNRoute projects,
    \item loading and processing YouTube traffic traces from the public dataset,
    \item generating the datasets varied by topologies and flow parameters,
    \item calculating metric values in all the datasets.    
\end{itemize}

We implemented all code in Python, utilizing scientific computing libraries including \emph{pandas}, \emph{numpy}, and \emph{networkx}. For the Growth \ac{GFU} metric, we formulated each iteration of the algorithm that determines the admissible $\alpha^{growth}$ vector as a \ac{LP} problem and solved it using the IBM ILOG CPLEX solver.

All metrics incorporated the adjustable $\alpha$ parameter described in Section~\ref{sec:load-impact-metrics}. In the experiments, we examined four distinct values of $\alpha$: $0.10$, $0.15$, $0.20$, and $0.25$. 

\subsection{Results}

Our evaluation calculated metric values for each experimental batch, applying the discussed metrics to network traffic data adapted to scenarios with chosen $\alpha$ parameter values. Each batch sequence progressively increased standard deviation and mean bandwidth of the added traffic, introducing higher traffic peaks. We analyzed how these systematic variations affected each metric's behavior. For practical reasons, we include only the most illustrative subset of graphs that highlight key findings. This selective approach is necessary, as presenting all 11 metrics across four $\alpha$ values would require 44 graphs --- exceeding practical space constraints. Instead, we offer a curated selection of graphs, accompanied by thorough commentary, to depict key observations and enable essential comparisons between metrics under the most important conditions.

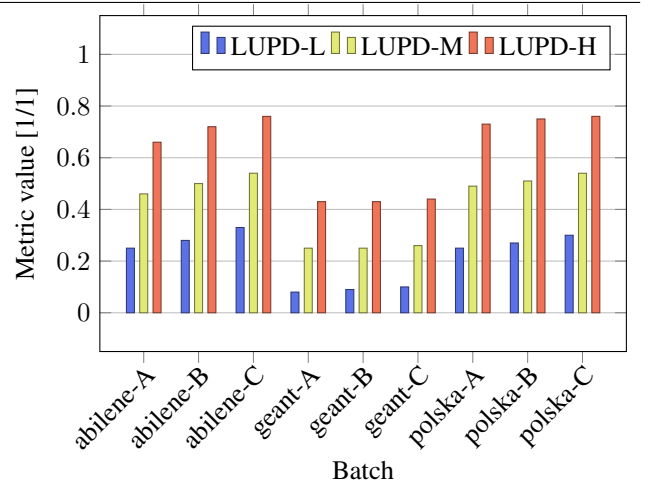
\begin{figure}[t!]
\centering
\begin{tikzpicture}
  \begin{axis}[
    table/col sep=comma,
    ybar,
    height=6cm,
    width=\linewidth,
    ymajorgrids, yminorgrids,
    ytick distance=0.2,
    ymin=-0.15, ymax=1.15,
    symbolic x coords={abilene-a, abilene-b, abilene-c, geant-a, geant-b, geant-c, polska-a, polska-b, polska-c},
    xtick={abilene-a, abilene-b, abilene-c, geant-a, geant-b, geant-c, polska-a, polska-b, polska-c},
    xticklabels={abilene-A, abilene-B, abilene-C, geant-A, geant-B, geant-C, polska-A, polska-B, polska-C},
    xticklabel style={rotate=45, anchor=north east, inner sep=0mm},
    bar width=3pt,
    legend entries={LUPD-L, LUPD-M, LUPD-H},
    legend columns=3,
    legend pos=north east,
    xlabel=Batch,
    ylabel={Metric value [1/1]},
    ]
    
    \addplot+[lcolor,draw=lcolor!50!black]
        table[x=batch,y=p_saturation_l] {flow_impact_results_std_a010.csv}; 
    \addplot+[mcolor,draw=mcolor!50!black]
        table[x=batch,y=p_saturation_m] {flow_impact_results_std_a010.csv};
    \addplot+[hcolor,draw=hcolor!50!black]
        table[x=batch,y=p_saturation_h] {flow_impact_results_std_a010.csv};
        
  \end{axis}
\end{tikzpicture}
\caption{\acf{LUPD} ($\alpha = 0.10$) \protect\linebreak Consistent increase in flow impact with bandwidth}
\label{fig:lupd-a010}
\end{figure}

The \ac{LUPD} metric demonstrates a consistent value increase with rising bandwidth across all batches, independent of traffic standard deviation and $\alpha$ parameter settings. Results for batches with identical standard deviation consistently differ by approximately $0.20$ in metric value. Fig.~\ref{fig:lupd-a010} illustrates these results for $\alpha = 0.10$. Specifically, the \emph{geant} topology consistently shows lower flow impact values, while both \emph{abilene} and \emph{polska} topologies demonstrate significantly higher impact measurements. This stems from the varied network link capacities and flow bandwidths added in particular scenarios. Additionally, increasing the standard deviation consistently produces slightly elevated flow impact measurements. Interestingly, the metric maintains consistent behavior across all $\alpha$ values, demonstrating robustness to parameter variations.

\begin{figure}[t!]
\centering
\begin{tikzpicture}
  \begin{axis}[
    table/col sep=comma,
    ybar,
    height=6cm,
    width=\linewidth,
    ymajorgrids, yminorgrids,
    ytick distance=0.2,
    ymin=-0.15, ymax=1.15,
    symbolic x coords={abilene-a, abilene-b, abilene-c, geant-a, geant-b, geant-c, polska-a, polska-b, polska-c},
    xtick={abilene-a, abilene-b, abilene-c, geant-a, geant-b, geant-c, polska-a, polska-b, polska-c},
    xticklabels={abilene-A, abilene-B, abilene-C, geant-A, geant-B, geant-C, polska-A, polska-B, polska-C},
    xticklabel style={rotate=45, anchor=north east, inner sep=0mm},
    bar width=3pt,
    legend entries={TLUSSD-L, TLUSSD-M, TLUSSD-H},
    xlabel=Batch,
    ylabel={Metric value [1/1]}
    ]
    
    \addplot+[lcolor,draw=lcolor!50!black]
        table[x=batch,y=cdf_saturation_l] {flow_impact_results_std_a010.csv}; 
    \addplot+[mcolor,draw=mcolor!50!black]
        table[x=batch,y=cdf_saturation_m] {flow_impact_results_std_a010.csv};
    \addplot+[hcolor,draw=hcolor!50!black]
        table[x=batch,y=cdf_saturation_h] {flow_impact_results_std_a010.csv};
        
  \end{axis}
\end{tikzpicture}
\caption{\acf{TLUSSD} ($\alpha = 0.10$) \protect\linebreak Minimal indicated flow impact for low-bandwidth cases due to low $\alpha$ value}
\label{fig:tlussd-a010}
\end{figure}
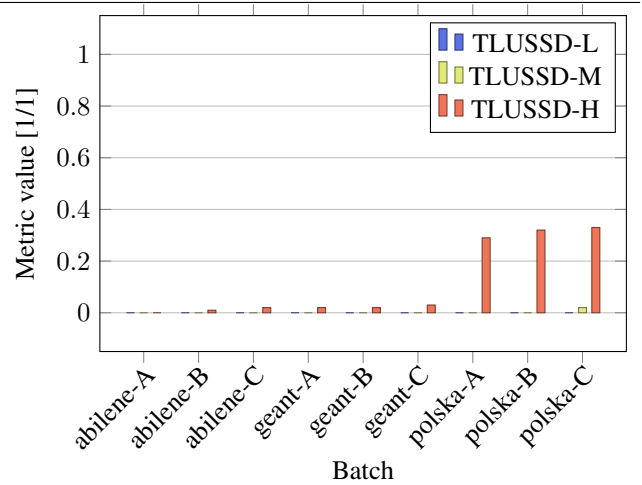

The \ac{TLUSSD} metric produces minimal values for L and M batches with low $\alpha$ values, indicating that traffic rarely exceeds the defined thresholds under these conditions. In contrast, H batches produce substantially higher values, particularly in the \emph{polska} topology, due to rapid link saturation in capacity-constrained networks. Higher $\alpha$ values establish more stringent thresholds, magnifying these differences and causing H batches to consistently register significantly higher values than both L and M batches, as shown in Fig.~\ref{fig:tlussd-a025}. Notably, the metric reveals negligible network impact across all L batches, confirming that the introduction of low-bandwidth flows fails to trigger significant threshold violations. For specific configurations (M and L batches with $\alpha$ values of $0.10$ and $0.15$), the metric registers precisely zero, indicating no noticeable flow impact, as illustrated in Fig.~\ref{fig:tlussd-a010}.

\begin{figure}[t]
\centering
\begin{tikzpicture}
  \begin{axis}[
    table/col sep=comma,
    ybar,
    height=6cm,
    width=\linewidth,
    ymajorgrids, yminorgrids,
    ytick distance=0.2,
    ymin=-0.15, ymax=1.15,
    symbolic x coords={abilene-a, abilene-b, abilene-c, geant-a, geant-b, geant-c, polska-a, polska-b, polska-c},
    xtick={abilene-a, abilene-b, abilene-c, geant-a, geant-b, geant-c, polska-a, polska-b, polska-c},
    xticklabels={abilene-A, abilene-B, abilene-C, geant-A, geant-B, geant-C, polska-A, polska-B, polska-C},
    xticklabel style={rotate=45, anchor=north east, inner sep=0mm},
    bar width=3pt,
    legend entries={TLUSSD-L, TLUSSD-M, TLUSSD-H},
    xlabel=Batch,
    ylabel={Metric value [1/1]},
    legend pos=north west
    ]
    
    \addplot+[lcolor,draw=lcolor!50!black]
        table[x=batch,y=cdf_saturation_l] {flow_impact_results_std_a025.csv}; 
    \addplot+[mcolor,draw=mcolor!50!black]
        table[x=batch,y=cdf_saturation_m] {flow_impact_results_std_a025.csv};
    \addplot+[hcolor,draw=hcolor!50!black]
        table[x=batch,y=cdf_saturation_h] {flow_impact_results_std_a025.csv};
        
  \end{axis}
\end{tikzpicture}
\caption{\acf{TLUSSD} ($\alpha = 0.25$) \protect\linebreak The metric is sensitive to link saturation, showing higher values in the \emph{polska} topology under high-load conditions}
\label{fig:tlussd-a025}
\end{figure}
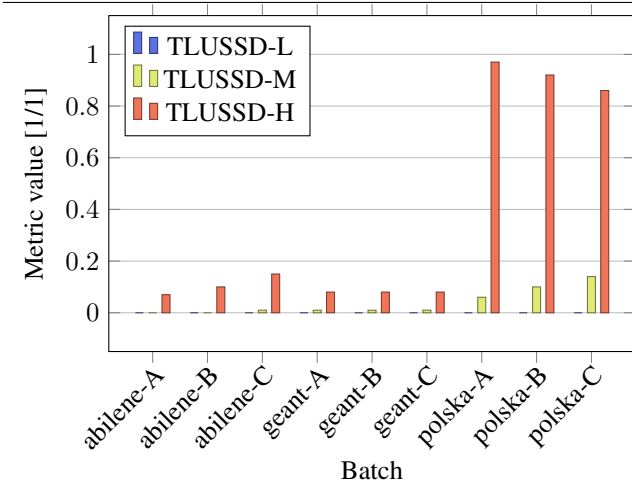

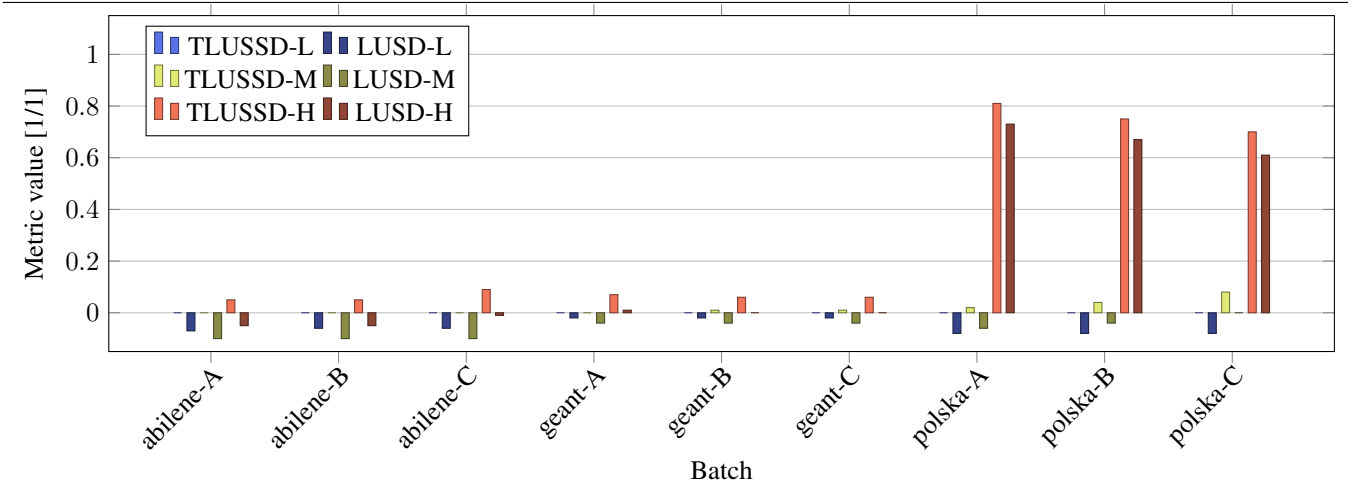
\begin{figure*}[ht!]
\centering
\begin{tikzpicture}
  \begin{axis}[
    table/col sep=comma,
    ybar,
    width=\textwidth,
    height=6cm,
    ymajorgrids, yminorgrids,
    ytick distance=0.2,
    ymin=-0.15, ymax=1.15,
    symbolic x coords={abilene-a, abilene-b, abilene-c, geant-a, geant-b, geant-c, polska-a, polska-b, polska-c},
    xtick={abilene-a, abilene-b, abilene-c, geant-a, geant-b, geant-c, polska-a, polska-b, polska-c},
    xticklabels={abilene-A, abilene-B, abilene-C, geant-A, geant-B, geant-C, polska-A, polska-B, polska-C},
    xticklabel style={rotate=45, anchor=north east, inner sep=0mm},
    bar width=3pt,
    legend entries={TLUSSD-L, LUSD-L, TLUSSD-M, LUSD-M, TLUSSD-H, LUSD-H},
    xlabel=Batch,
    ylabel={Metric value [1/1]},
    legend pos=north west,
    legend columns=2
    ]
    
    \addplot+[lcolor,draw=lcolor!50!black]
        table[x=batch,y=cdf_saturation_l] {flow_impact_results_std_a020.csv}; 
    \addplot+[lcolor!60!black,draw=lcolor!30!black]
        table[x=batch,y=uscore_saturation_l] {flow_impact_results_std_a020.csv};
    
    \addplot+[mcolor,draw=mcolor!50!black]
        table[x=batch,y=cdf_saturation_m] {flow_impact_results_std_a020.csv};
    \addplot+[mcolor!60!black,draw=mcolor!30!black]
        table[x=batch,y=uscore_saturation_m] {flow_impact_results_std_a020.csv};

    \addplot+[hcolor,draw=hcolor!50!black]
        table[x=batch,y=cdf_saturation_h] {flow_impact_results_std_a020.csv};
    \addplot+[hcolor!60!black,draw=hcolor!30!black]
        table[x=batch,y=uscore_saturation_h] {flow_impact_results_std_a020.csv};
        
  \end{axis}
\end{tikzpicture}
\caption{\acf{TLUSSD}, \acf{LUSD} ($\alpha = 0.20$) \protect\linebreak Both metrics yield similar results, however \acs{LUSD} values consider both under- and overutilization}
\label{fig:tlussd-lusd-a020}
\end{figure*}

\begin{figure}[t]
\centering
\begin{tikzpicture}
  \begin{axis}[
    table/col sep=comma,
    ybar,
    width=\linewidth,
    height=6cm,
    ymajorgrids, yminorgrids,
    ytick distance=0.2,
    ymin=-0.15, ymax=1.15,
    symbolic x coords={abilene-a, abilene-b, abilene-c, geant-a, geant-b, geant-c, polska-a, polska-b, polska-c},
    xtick={abilene-a, abilene-b, abilene-c, geant-a, geant-b, geant-c, polska-a, polska-b, polska-c},
    xticklabels={abilene-A, abilene-B, abilene-C, geant-A, geant-B, geant-C, polska-A, polska-B, polska-C},
    xticklabel style={rotate=45, anchor=north east, inner sep=0mm},
    bar width=3pt,
    legend entries={LUSD-L, LUSD-M, LUSD-H},
    xlabel=Batch,
    ylabel={Metric value [1/1]},
    ]
    
    \addplot+[lcolor,draw=lcolor!50!black]
        table[x=batch,y=uscore_saturation_l] {flow_impact_results_std_a015.csv}; 
    \addplot+[mcolor,draw=mcolor!50!black]
        table[x=batch,y=uscore_saturation_m] {flow_impact_results_std_a015.csv};
    \addplot+[hcolor,draw=hcolor!50!black]
        table[x=batch,y=uscore_saturation_h] {flow_impact_results_std_a015.csv};
        
  \end{axis}
\end{tikzpicture}
\caption{\acf{LUSD} ($\alpha = 0.15$) \protect\linebreak The metric indicates improved resource utilization with negative values}
\label{fig:lusd-a015}
\end{figure}
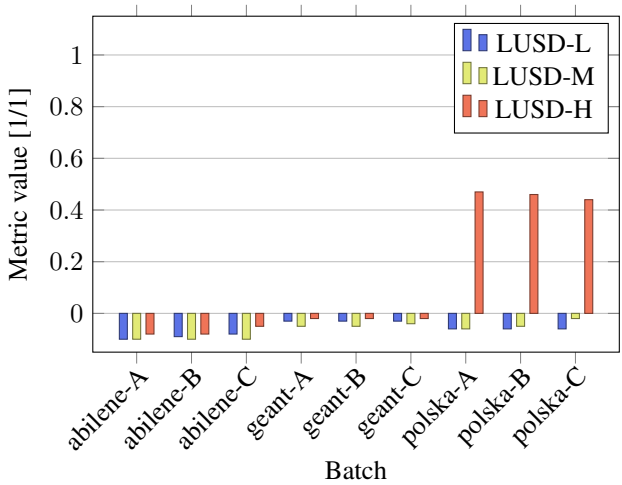

The \ac{LUSD} metric exhibits behavior similar to \ac{TLUSSD}, particularly for H batches in the \emph{polska} topology across all $\alpha$ values. Fig.~\ref{fig:tlussd-lusd-a020} presents a comparative analysis of both metrics for $\alpha = 0.20$. Conversely, L and M batches exhibit negative values ranging from $0$ to $-0.10$, as shown in Fig.~\ref{fig:lusd-a015}. This characteristic stems from the metric's ability to evaluate both under- and overutilization simultaneously, extending beyond the capabilities implemented in \ac{TLUSSD}. Negative metric values signify an improvement in overall resource efficiency, indicating scenarios where previously underutilized network resources achieve more optimal utilization through the introduction of additional flows, without approaching capacity limitations.

The \ac{MLUPD} metric produces results that generally align with \ac{LUPD} behavior. However, a distinctive pattern emerges in the \emph{geant} topology: \ac{MLUPD} registers higher values for L and M batches compared to \ac{LUPD}, while exhibiting lower values for H batches. The most significant divergence occurs within the \emph{geant} topology, while the other topologies exhibit greater consistency between the two metrics. Fig.~\ref{fig:lupd-mlupd-a020} illustrates this behavior for $\alpha = 0.20$. These findings indicate that the averaging process in mean link utilization calculations can mask local under- or overutilization events occurring on individual links within a network topology. Furthermore, in networks with homogeneous load distribution and uniform link capacities, such as the \emph{polska} topology, \ac{MLUPD} produces results that closely correspond with \ac{LUPD} measurements. In the opposite case, as seen in the instance of the \emph{geant} topology, a greater number of divergences between the two metrics are noted.

\begin{figure*}
\centering
\begin{tikzpicture}
  \begin{axis}[
    table/col sep=comma,
    ybar,
    width=\textwidth,
    height=6cm,
    ymajorgrids, yminorgrids,
    ytick distance=0.2,
    ymin=-0.15, ymax=1.15,
    symbolic x coords={abilene-a, abilene-b, abilene-c, geant-a, geant-b, geant-c, polska-a, polska-b, polska-c},
    xtick={abilene-a, abilene-b, abilene-c, geant-a, geant-b, geant-c, polska-a, polska-b, polska-c},
    xticklabels={abilene-A, abilene-B, abilene-C, geant-A, geant-B, geant-C, polska-A, polska-B, polska-C},
    xticklabel style={rotate=45, anchor=north east, inner sep=0mm},
    bar width=3pt,
    legend entries={LUPD-L, MLUPD-L, LUPD-M, MLUPD-M, LUPD-H, MLUPD-H},
    xlabel=Batch,
    ylabel={Metric value [1/1]},
    legend pos=north west,
    legend columns=2
    ]
    
    \addplot+[lcolor,draw=lcolor!50!black]
        table[x=batch,y=p_saturation_l] {flow_impact_results_std_a020.csv}; 
    \addplot+[lcolor!60!black,draw=lcolor!30!black]
        table[x=batch,y=p_mean_saturation_l] {flow_impact_results_std_a020.csv};
    
    \addplot+[mcolor,draw=mcolor!50!black]
        table[x=batch,y=p_saturation_m] {flow_impact_results_std_a020.csv};
    \addplot+[mcolor!60!black,draw=mcolor!30!black]
        table[x=batch,y=p_mean_saturation_m] {flow_impact_results_std_a020.csv};

    \addplot+[hcolor,draw=hcolor!50!black]
        table[x=batch,y=p_saturation_h] {flow_impact_results_std_a020.csv};
    \addplot+[hcolor!60!black,draw=hcolor!30!black]
        table[x=batch,y=p_mean_saturation_h] {flow_impact_results_std_a020.csv};
        
  \end{axis}
\end{tikzpicture}
\caption{\acf{LUPD}, \acf{MLUPD} ($\alpha = 0.20$) \protect\linebreak The \acs{MLUPD} metric based on mean link utilization may overlook under- or overutilization occurring on single links}
\label{fig:lupd-mlupd-a020}
\end{figure*}

For the reasons discussed above, the \ac{TMLUSSD} metric reveals minimal detectable flow impact for L and M batches, as illustrated in Fig.~\ref{fig:tmlussd-a010}. This limited sensitivity is further influenced by the selection of low $\alpha$ values. In contrast, the metric produces maximum values (up to $1$) in the \emph{polska} topology with H batches where widespread link saturation occurs, as shown in Fig.~\ref{fig:tmlussd-a025}. This effect is amplified with increasing $\alpha$ values. The exceptionally high value recorded for the \emph{polska-A-H} scenario indicates that all measured samples exceeded the seemingly low threshold. Analysis reveals that \ac{TMLUSSD} demonstrates reduced sensitivity to low utilization, while significantly amplifying the measurement of high utilization. This measurement bias causes the metric to emphasize high utilization disproportionately in various network scenarios. Fig.~\ref{fig:tlussd-tmlussd-a025} provides a comparative analysis of \ac{TLUSSD} and \ac{TMLUSSD} values for $\alpha = 0.25$, further demonstrating this measurement characteristic.

\begin{figure}[ht]
\centering
\begin{tikzpicture}
  \begin{axis}[
    table/col sep=comma,
    ybar,
    width=\linewidth,
    height=6cm,
    ymajorgrids, yminorgrids,
    ytick distance=0.2,
    ymin=-0.15, ymax=1.15,
    symbolic x coords={abilene-a, abilene-b, abilene-c, geant-a, geant-b, geant-c, polska-a, polska-b, polska-c},
    xtick={abilene-a, abilene-b, abilene-c, geant-a, geant-b, geant-c, polska-a, polska-b, polska-c},
    xticklabels={abilene-A, abilene-B, abilene-C, geant-A, geant-B, geant-C, polska-A, polska-B, polska-C},
    xticklabel style={rotate=45, anchor=north east, inner sep=0mm},
    bar width=3pt,
    legend entries={TMLUSSD-L, TMLUSSD-M, TMLUSSD-H},
    xlabel=Batch,
    ylabel={Metric value [1/1]}
    ]
    
    \addplot+[lcolor,draw=lcolor!50!black]
        table[x=batch,y=cdf_mean_saturation_l] {flow_impact_results_std_a010.csv}; 
    \addplot+[mcolor,draw=mcolor!50!black]
        table[x=batch,y=cdf_mean_saturation_m] {flow_impact_results_std_a010.csv};
    \addplot+[hcolor,draw=hcolor!50!black]
        table[x=batch,y=cdf_mean_saturation_h] {flow_impact_results_std_a010.csv};
        
  \end{axis}
\end{tikzpicture}
\caption{\acf{TMLUSSD} ($\alpha = 0.10$) \protect\linebreak Due to the application of mean link utilization, flow impact is indicated only with the highest bandwidth level and most of network links saturated}
\label{fig:tmlussd-a010}
\end{figure}
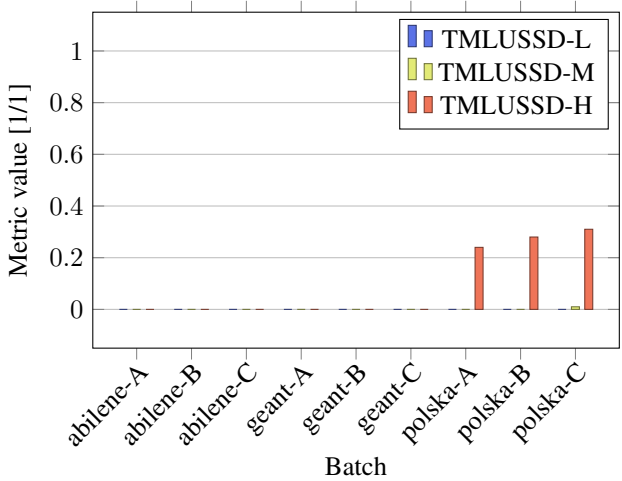

\begin{figure}[ht]
\centering
\begin{tikzpicture}
  \begin{axis}[
    table/col sep=comma,
    ybar,
    width=\linewidth,
    height=6cm,
    ymajorgrids, yminorgrids,
    ytick distance=0.2,
    ymin=-0.15, ymax=1.15,
    symbolic x coords={abilene-a, abilene-b, abilene-c, geant-a, geant-b, geant-c, polska-a, polska-b, polska-c},
    xtick={abilene-a, abilene-b, abilene-c, geant-a, geant-b, geant-c, polska-a, polska-b, polska-c},
    xticklabels={abilene-A, abilene-B, abilene-C, geant-A, geant-B, geant-C, polska-A, polska-B, polska-C},
    xticklabel style={rotate=45, anchor=north east, inner sep=0mm},
    bar width=3pt,
    legend entries={TMLUSSD-L, TMLUSSD-M, TMLUSSD-H},
    xlabel=Batch,
    ylabel={Metric value [1/1]},
    legend pos=north west
    ]
    
    \addplot+[lcolor,draw=lcolor!50!black]
        table[x=batch,y=cdf_mean_saturation_l] {flow_impact_results_std_a025.csv}; 
    \addplot+[mcolor,draw=mcolor!50!black]
        table[x=batch,y=cdf_mean_saturation_m] {flow_impact_results_std_a025.csv};
    \addplot+[hcolor,draw=hcolor!50!black]
        table[x=batch,y=cdf_mean_saturation_h] {flow_impact_results_std_a025.csv};
        
  \end{axis}
\end{tikzpicture}
\caption{\acf{TMLUSSD} ($\alpha = 0.25$) \protect\linebreak The metric value of 1 indicates that all the samples exceeded the threshold}
\label{fig:tmlussd-a025}
\end{figure}
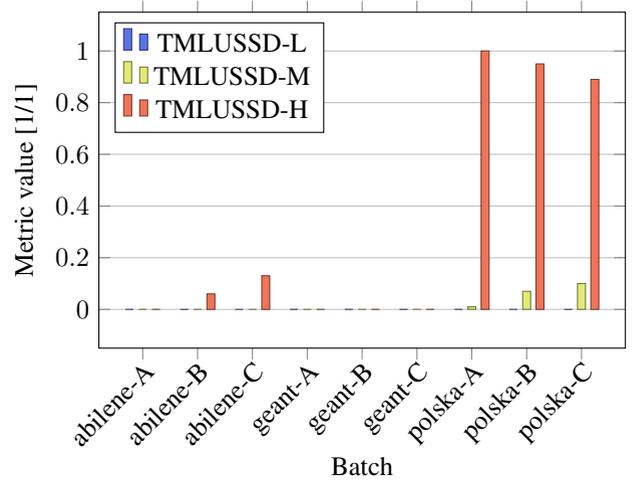

\begin{figure*}
\centering
\begin{tikzpicture}
  \begin{axis}[
    table/col sep=comma,
    ybar,
    width=\textwidth,
    height=6cm,
    ymajorgrids, yminorgrids,
    ytick distance=0.2,
    ymin=-0.15, ymax=1.15,
    symbolic x coords={abilene-a, abilene-b, abilene-c, geant-a, geant-b, geant-c, polska-a, polska-b, polska-c},
    xtick={abilene-a, abilene-b, abilene-c, geant-a, geant-b, geant-c, polska-a, polska-b, polska-c},
    xticklabels={abilene-A, abilene-B, abilene-C, geant-A, geant-B, geant-C, polska-A, polska-B, polska-C},
    xticklabel style={rotate=45, anchor=north east, inner sep=0mm},
    bar width=3pt,
    legend entries={TLUSSD-L, TMLUSSD-L, TLUSSD-M, TMLUSSD-M, TLUSSD-H, TMLUSSD-H},
    xlabel=Batch,
    ylabel={Metric value [1/1]},
    legend pos=north west,
    legend columns=2
    ]
    
    \addplot+[lcolor,draw=lcolor!50!black]
        table[x=batch,y=cdf_saturation_l] {flow_impact_results_std_a025.csv}; 
    \addplot+[lcolor!60!black,draw=lcolor!30!black]
        table[x=batch,y=cdf_mean_saturation_l] {flow_impact_results_std_a025.csv};
    
    \addplot+[mcolor,draw=mcolor!50!black]
        table[x=batch,y=cdf_saturation_m] {flow_impact_results_std_a025.csv};
    \addplot+[mcolor!60!black,draw=mcolor!30!black]
        table[x=batch,y=cdf_mean_saturation_m] {flow_impact_results_std_a025.csv};

    \addplot+[hcolor,draw=hcolor!50!black]
        table[x=batch,y=cdf_saturation_h] {flow_impact_results_std_a025.csv};
    \addplot+[hcolor!60!black,draw=hcolor!30!black]
        table[x=batch,y=cdf_mean_saturation_h] {flow_impact_results_std_a025.csv};
        
  \end{axis}
\end{tikzpicture}
\caption{\acf{TLUSSD}, \acf{TMLUSSD} ($\alpha = 0.25$) \protect\linebreak \acs{TMLUSSD} demonstrates greater sensitivity to flow impact in high-load scenarios compared to \acs{TLUSSD}, contrary to low-load scenarios}
\label{fig:tlussd-tmlussd-a025}
\end{figure*}

The \ac{GGPD} and \ac{RGPD} metrics exhibit remarkable consistency with each other, with maximum differences not exceeding $0.02$ across all test conditions. Fig.~\ref{fig:ggpd-rgpd-a025} illustrates this close correlation between both metrics for $\alpha = 0.25$. Both metrics produce results that align closely with \ac{LUPD} and \ac{MLUPD} in the \emph{abilene} and \emph{polska} topologies, while registering substantially lower values in the \emph{geant} topology. In the \emph{geant} topology specifically, these metrics register values $1.50$ to $5$ times lower than their \ac{LUPD} and \ac{MLUPD} counterparts. \ac{GGPD} and \ac{RGPD} reveal fundamentally different flow impact assessments in certain scenarios compared to other metrics because the underlying \ac{GFU} calculation incorporates additional network factors, particularly the presence and distribution of concurrent flows. Consequently, in overprovisioned networks like the \emph{geant} topology, these metrics accurately identify substantial remaining capacity to accommodate additional traffic --- a capability not reflected in the standard link-utilization metrics. Fig.~\ref{fig:lupd-mlupd-ggpd-rgpd-a010} provides a comparison of all four metrics (\ac{LUPD}, \ac{MLUPD}, \ac{GGPD}, and \ac{RGPD}) for $\alpha = 0.10$.

\begin{figure*}
\centering
\begin{tikzpicture}
  \begin{axis}[
    table/col sep=comma,
    ybar,
    width=\textwidth,
    height=6cm,
    ymajorgrids, yminorgrids,
    ytick distance=0.2,
    ymin=-0.15, ymax=1.15,
    symbolic x coords={abilene-a, abilene-b, abilene-c, geant-a, geant-b, geant-c, polska-a, polska-b, polska-c},
    xtick={abilene-a, abilene-b, abilene-c, geant-a, geant-b, geant-c, polska-a, polska-b, polska-c},
    xticklabels={abilene-A, abilene-B, abilene-C, geant-A, geant-B, geant-C, polska-A, polska-B, polska-C},
    xticklabel style={rotate=45, anchor=north east, inner sep=0mm},
    bar width=3pt,
    legend entries={GGPD-L, RGPD-L, GGPD-M, GGPD-M, GGPD-H, RGPD-H},
    xlabel=Batch,
    ylabel={Metric value [1/1]},
    legend pos=north west,
    legend columns=2
    ]
    
    \addplot+[lcolor,draw=lcolor!50!black]
        table[x=batch,y=p_gfu_growth_l] {flow_impact_results_std_a025.csv};
    \addplot+[lcolor!60!black,draw=lcolor!30!black]
        table[x=batch,y=p_gfu_risk_l] {flow_impact_results_std_a025.csv};
    
    \addplot+[mcolor,draw=mcolor!50!black]
        table[x=batch,y=p_gfu_growth_m] {flow_impact_results_std_a025.csv};
    \addplot+[mcolor!60!black,draw=mcolor!30!black]
        table[x=batch,y=p_gfu_risk_m] {flow_impact_results_std_a025.csv};
        
    \addplot+[hcolor,draw=hcolor!50!black]
        table[x=batch,y=p_gfu_growth_h] {flow_impact_results_std_a025.csv};
    \addplot+[hcolor!60!black,draw=hcolor!30!black]
        table[x=batch,y=p_gfu_risk_h] {flow_impact_results_std_a025.csv};
        
  \end{axis}
\end{tikzpicture}
\caption{\acf{GGPD}, \acf{RGPD} ($\alpha = 0.25$) \protect\linebreak Values of \acs{GGPD} and \acs{RGPD} are closely aligned, exhibiting only minimal differences}
\label{fig:ggpd-rgpd-a025}
\end{figure*}

\begin{figure*}
\centering
\begin{tikzpicture}
  \begin{axis}[
    table/col sep=comma,
    ybar,
    width=\textwidth,
    height=6cm,
    ymajorgrids, yminorgrids,
    ytick distance=0.2,
    ymin=-0.15, ymax=1.15,
    symbolic x coords={abilene-a, abilene-b, abilene-c, geant-a, geant-b, geant-c, polska-a, polska-b, polska-c},
    xtick={abilene-a, abilene-b, abilene-c, geant-a, geant-b, geant-c, polska-a, polska-b, polska-c},
    xticklabels={abilene-A, abilene-B, abilene-C, geant-A, geant-B, geant-C, polska-A, polska-B, polska-C},
    xticklabel style={rotate=45, anchor=north east, inner sep=0mm},
    bar width=1.5pt,
    legend entries={LUPD-L, MLUPD-L, GGPD-L, RGPD-L, LUPD-M, MLUPD-M, GGPD-M, GGPD-M, LUPD-H, MLUPD-H, GGPD-H, RGPD-H},
    xlabel=Batch,
    ylabel={Metric value [1/1]},
    legend columns=4,
    legend style={anchor=center,at={(0.75, 0.9)}}
    ]
    
    \addplot+[lcolor,draw=lcolor!80!black]
        table[x=batch,y=p_saturation_l] {flow_impact_results_std_a010.csv}; 
    \addplot+[lcolor!80!black,draw=lcolor!60!black]
        table[x=batch,y=p_mean_saturation_l] {flow_impact_results_std_a010.csv}; 
    \addplot+[lcolor!60!black,draw=lcolor!40!black]
        table[x=batch,y=p_gfu_growth_l] {flow_impact_results_std_a010.csv};
    \addplot+[lcolor!40!black,draw=lcolor!20!black]
        table[x=batch,y=p_gfu_risk_l] {flow_impact_results_std_a010.csv};
    
    \addplot+[mcolor,draw=mcolor!80!black]
        table[x=batch,y=p_saturation_m] {flow_impact_results_std_a010.csv};
    \addplot+[mcolor!80!black,draw=mcolor!60!black]
        table[x=batch,y=p_mean_saturation_m] {flow_impact_results_std_a010.csv};
    \addplot+[mcolor!60!black,draw=mcolor!40!black]
        table[x=batch,y=p_gfu_growth_m] {flow_impact_results_std_a010.csv};
    \addplot+[mcolor!40!black,draw=mcolor!20!black]
        table[x=batch,y=p_gfu_risk_m] {flow_impact_results_std_a010.csv};
        
    \addplot+[hcolor,draw=hcolor!80!black]
        table[x=batch,y=p_saturation_h] {flow_impact_results_std_a010.csv};
    \addplot+[hcolor!80!black,draw=hcolor!60!black]
        table[x=batch,y=p_mean_saturation_h] {flow_impact_results_std_a010.csv};
    \addplot+[hcolor!60!black,draw=hcolor!40!black]
        table[x=batch,y=p_gfu_growth_h] {flow_impact_results_std_a010.csv};
    \addplot+[hcolor!40!black,draw=hcolor!20!black]
        table[x=batch,y=p_gfu_risk_h] {flow_impact_results_std_a010.csv};
        
  \end{axis}
\end{tikzpicture}
\caption{\acf{LUPD}, \acf{MLUPD}, \acf{GGPD}, \acf{RGPD} ($\alpha = 0.10$) \protect\linebreak \acs{GGPD} and \acs{RGPD} align with \acs{LUPD} and \acs{MLUPD} in the \emph{abilene} and \emph{polska} topologies but show lower values in \emph{geant}, taking into account factors like future traffic accommodation considered in \acs{GFU}}
\label{fig:lupd-mlupd-ggpd-rgpd-a010}
\end{figure*}

The \ac{TGGSSD} and \ac{TRGSSD} metrics typically produce values within $0.02$ of \ac{TLUSSD} measurements for the \emph{abilene} topology and for L-M batches in the \emph{geant} topology. Significant divergence between these metrics emerges in the \emph{geant} topology with H batches, where \ac{TLUSSD} consistently registers notably higher values than either \ac{TGGSSD} or \ac{TRGSSD} --- a disparity that increases progressively with higher $\alpha$ parameter settings. In the \emph{polska} topology, a similar pattern emerges for L and M batches, with all three metrics producing values within a $0.02$ range. However, for H batches, this pattern reverses, with \ac{TGGSSD} and \ac{TRGSSD} exceeding \ac{TLUSSD} values by up to $0.15$. Fig.~\ref{fig:tlussd-tggssd-trgssd-a020} illustrates the comparative behavior of all three metrics for scenarios with $\alpha = 0.20$.

\begin{figure*}
\centering
\begin{tikzpicture}
  \begin{axis}[
    table/col sep=comma,
    ybar,
    width=\textwidth,
    height=6cm,
    ymajorgrids, yminorgrids,
    ytick distance=0.2,
    ymin=-0.15, ymax=1.15,
    symbolic x coords={abilene-a, abilene-b, abilene-c, geant-a, geant-b, geant-c, polska-a, polska-b, polska-c},
    xtick={abilene-a, abilene-b, abilene-c, geant-a, geant-b, geant-c, polska-a, polska-b, polska-c},
    xticklabels={abilene-A, abilene-B, abilene-C, geant-A, geant-B, geant-C, polska-A, polska-B, polska-C},
    xticklabel style={rotate=45, anchor=north east, inner sep=0mm},
    bar width=3pt,
    legend entries={TLUSSD-L, TGGSSD-L, TRGSSD-L,  LUSSD-M, TGGSSD-M, TRGSSD-M, TLUSSD-H, TGGSSD-H, TRGSSD-H},
    xlabel=Batch,
    ylabel={Metric value [1/1]},
    legend pos=north west,
    legend columns=3
    ]
    
    \addplot+[lcolor,draw=lcolor!50!black]
        table[x=batch,y=cdf_saturation_l] {flow_impact_results_std_a020.csv}; 
    \addplot+[lcolor!80!black,draw=lcolor!40!black]
        table[x=batch,y=cdf_gfu_growth_l] {flow_impact_results_std_a020.csv};
    \addplot+[lcolor!60!black,draw=lcolor!30!black]
        table[x=batch,y=cdf_gfu_risk_l] {flow_impact_results_std_a020.csv}; 
    
    \addplot+[mcolor,draw=mcolor!50!black]
        table[x=batch,y=cdf_saturation_m] {flow_impact_results_std_a020.csv};
    \addplot+[mcolor!80!black,draw=mcolor!40!black]
        table[x=batch,y=cdf_gfu_growth_m] {flow_impact_results_std_a020.csv};
    \addplot+[mcolor!60!black,draw=mcolor!30!black]
        table[x=batch,y=cdf_gfu_risk_m] {flow_impact_results_std_a020.csv};
        
    \addplot+[hcolor,draw=hcolor!50!black]
        table[x=batch,y=cdf_saturation_h] {flow_impact_results_std_a020.csv};
    \addplot+[hcolor!80!black,draw=hcolor!40!black]
        table[x=batch,y=cdf_gfu_growth_h] {flow_impact_results_std_a020.csv};
    \addplot+[hcolor!60!black,draw=hcolor!30!black]
        table[x=batch,y=cdf_gfu_risk_h] {flow_impact_results_std_a020.csv};
        
  \end{axis}
\end{tikzpicture}
\caption{\acf{TLUSSD}, \acf{TGGSSD}, \acf{TRGSSD} ($\alpha = 0.20$) \protect\linebreak \acs{TGGSSD} and \acs{TRGSSD} align with \acs{TLUSSD} in most cases, with slight difference in high-load batches}
\label{fig:tlussd-tggssd-trgssd-a020}
\end{figure*}

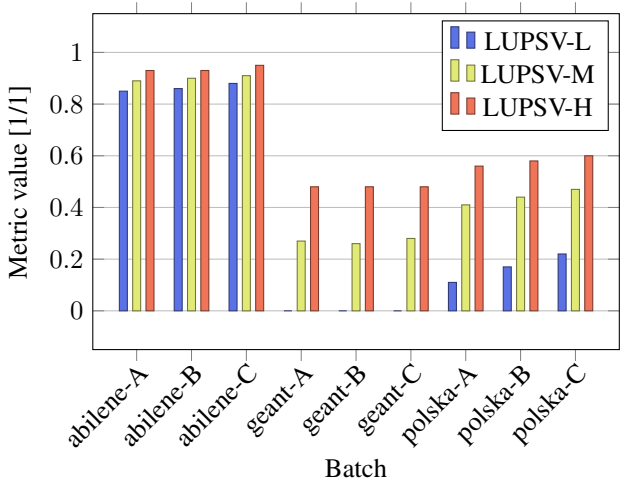
\begin{figure}[ht]
\centering
\begin{tikzpicture}
  \begin{axis}[
    table/col sep=comma,
    ybar,
    width=\linewidth,
    height=6cm,
    ymajorgrids, yminorgrids,
    ytick distance=0.2,
    ymin=-0.15, ymax=1.15,
    symbolic x coords={abilene-a, abilene-b, abilene-c, geant-a, geant-b, geant-c, polska-a, polska-b, polska-c},
    xtick={abilene-a, abilene-b, abilene-c, geant-a, geant-b, geant-c, polska-a, polska-b, polska-c},
    xticklabels={abilene-A, abilene-B, abilene-C, geant-A, geant-B, geant-C, polska-A, polska-B, polska-C},
    xticklabel style={rotate=45, anchor=north east, inner sep=0mm},
    bar width=3pt,
    legend entries={LUPSV-L, LUPSV-M, LUPSV-H},
    xlabel=Batch,
    ylabel={Metric value [1/1]}
    ]
    
    \addplot+[lcolor,draw=lcolor!50!black]
        table[x=batch,y=shap_saturation_percentile_l] {flow_impact_results_std_a015.csv}; 
    \addplot+[mcolor,draw=mcolor!50!black]
        table[x=batch,y=shap_saturation_percentile_m] {flow_impact_results_std_a015.csv};
    \addplot+[hcolor,draw=hcolor!50!black]
        table[x=batch,y=shap_saturation_percentile_h] {flow_impact_results_std_a015.csv};
        
  \end{axis}
\end{tikzpicture}
\caption{\acf{LUPSV} ($\alpha = 0.15$) \protect\linebreak The metric considers absolute network load levels, resulting in different results compared to metrics like \acs{LUPD}, particularly in low-load batches in the \emph{geant} topology}
\label{fig:lupsv-a015}
\end{figure}

The \ac{LUPSV} metric consistently registers zero values for all L batches in the \emph{geant} topology, confirming the negligible impact of introduced flows on overall network traffic. The metric produces substantially higher values for the \emph{abilene} and \emph{geant} topologies (excluding L batches), while consistently yielding lower measurements for the \emph{polska} topology. This metric incorporates additional factors, revealing different flow impact profiles compared to the previously analyzed metrics. For example, while the \ac{LUPD} metric produces identical values of $0.23$ for both the \emph{abilene-A-L} and \emph{geant-C-M} scenarios at $\alpha = 0.15$, the \ac{LUPSV} metric reveals remarkable differences. Specifically, \ac{LUPSV} distinguishes these scenarios with substantially different values: $0.85$ for \emph{abilene-A-L} and $0.28$ for \emph{geant-C-M}, as illustrated in Fig.~\ref{fig:lupsv-a015}. This stems from the \ac{LUPSV} methodology, which evaluates absolute network load levels in both the initial state and after flow introduction, rather than merely measuring relative changes.

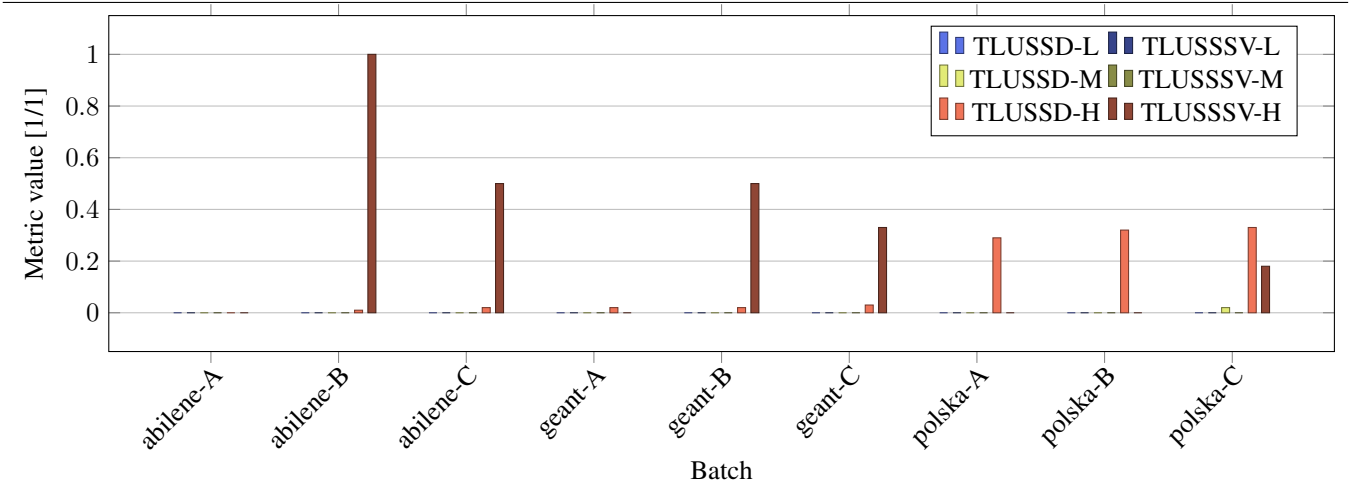
\begin{figure*}[ht]
\centering
\begin{tikzpicture}
  \begin{axis}[
    table/col sep=comma,
    ybar,
    width=\textwidth,
    height=6cm,
    ymajorgrids, yminorgrids,
    ymin=-0.15, ymax=1.15,
    ytick distance=0.2,
    symbolic x coords={abilene-a, abilene-b, abilene-c, geant-a, geant-b, geant-c, polska-a, polska-b, polska-c},
    xtick={abilene-a, abilene-b, abilene-c, geant-a, geant-b, geant-c, polska-a, polska-b, polska-c},
    xticklabels={abilene-A, abilene-B, abilene-C, geant-A, geant-B, geant-C, polska-A, polska-B, polska-C},
    xticklabel style={rotate=45, anchor=north east, inner sep=0mm},
    bar width=3pt,
    legend entries={TLUSSD-L, TLUSSSV-L, TLUSSD-M, TLUSSSV-M, TLUSSD-H, TLUSSSV-H},
    xlabel=Batch,
    ylabel={Metric value [1/1]},
    legend pos=north east,
    legend columns=2
    ]
    
    \addplot+[lcolor,draw=lcolor!50!black]
        table[x=batch,y=cdf_saturation_l] {flow_impact_results_std_a010.csv};
    \addplot+[lcolor!60!black,draw=lcolor!30!black]
        table[x=batch,y=shap_saturation_cdf_l] {flow_impact_results_std_a010.csv};
    
    \addplot+[mcolor,draw=mcolor!50!black]
        table[x=batch,y=cdf_saturation_m] {flow_impact_results_std_a010.csv};
    \addplot+[mcolor!60!black,draw=mcolor!30!black]
        table[x=batch,y=shap_saturation_cdf_m] {flow_impact_results_std_a010.csv};
        
    \addplot+[hcolor,draw=hcolor!50!black]
        table[x=batch,y=cdf_saturation_h] {flow_impact_results_std_a010.csv};
    \addplot+[hcolor!60!black,draw=hcolor!30!black]
        table[x=batch,y=shap_saturation_cdf_h] {flow_impact_results_std_a010.csv};
        
  \end{axis}
\end{tikzpicture}
\caption{\acf{TLUSSD}, \acf{TLUSSSV} ($\alpha = 0.10$) \protect\linebreak \acs{TLUSSSV} yields higher values than \acs{TLUSSD} in H batches for \emph{abilene} and \emph{geant}, but lower values in \emph{polska}, as it accounts for both the initial network state and the added flow}
\label{fig:tlussd-tlusssv-a010}
\end{figure*}

The \ac{TLUSSSV} metric frequently registers substantially higher values than \ac{TLUSSD}, particularly for H batches in both the \emph{abilene} and \emph{geant} topologies —-- a pattern consistent across all $\alpha$ parameter settings. Fig.~\ref{fig:tlussd-tlusssv-a010} illustrates this relationship between the metrics for $\alpha = 0.10$. Conversely, in the \emph{polska} topology with H batches, \ac{TLUSSSV} produces notably lower values than its \ac{TLUSSD} counterpart. This stems from the Shapley value methodology, which evaluates not only the initial network state but also incorporates the independent metric value of the added flow, providing a fundamentally different perspective on flow impact assessment.

\subsection{Summary}

The metrics presented in this study reveal consistent patterns across topologies and $\alpha$ values, providing diverse insights into flow impact on network state. The percentile-based \ac{LUPD} metric demonstrates high sensitivity, detecting even minor network state changes caused by the introduction of low-bandwidth flows. This responsiveness makes \ac{LUPD} particularly valuable for assessing immediate network state changes. The \ac{MLUPD} metric yields largely comparable results, though it frequently fails to detect localized overutilization or underutilization events on specific links. This limitation stems from its reliance on averaged link utilization values, which biases results toward the predominant state of the network infrastructure. This aggregation approach, however, provides computational advantages through reduced data storage and processing requirements compared to \ac{LUPD}.

The \ac{GGPD} and \ac{RGPD} metrics demonstrate similarly consistent performance characteristics. These metrics most closely align with \ac{LUPD} and \ac{MLUPD} behavior in the \emph{abilene} and \emph{polska} topologies. However, in the \emph{geant} topology scenarios, both metrics produce substantially lower values than their \ac{LUPD} or \ac{MLUPD} counterparts. This divergence stems from their flow-based methodology, which evaluates network impact in relation to the topology's capacity to accommodate additional flows. It should be noted that \ac{GGPD} calculation imposes substantial computational demands due to its input data requirements and algorithmic complexity. However, \ac{RGPD} provides a computationally efficient alternative while maintaining comparable capabilities.

The \ac{LUPSV} metric offers a different analytical perspective by quantifying the magnitude of network state modifications resulting from flow introduction. Its correlation with other percentile-based metrics varies. Specifically, in \emph{abilene} topology scenarios, \ac{LUPSV} consistently produces higher values than comparable metrics, while in \emph{polska} topology configurations, it generates systematically lower measurements. The \emph{geant} topology presents particularly interesting behavior, where \ac{LUPSV}'s relationship to other metrics varies within individual scenarios --- some low values are reduced to zero, while simultaneously producing significantly amplified measurements in other cases. This can be explained by the \ac{LUPSV}'s methodology, which evaluates network modifications by incorporating both initial network state and link capacity distributions --- different from the absolute difference calculations employed by other metrics.

Threshold-based metrics that quantify the proportion of samples exceeding predefined limits, including \ac{TLUSSD}, \ac{TMLUSSD}, \ac{TGGSSD}, and \ac{TRGSSD}, generally produce consistent results across test scenarios. These metrics frequently return zero values when traffic levels remain below saturation thresholds, even after additional flow introduction. When producing non-zero measurements, \ac{TGGSSD} typically registers higher values than other threshold-based metrics, particularly in the \emph{polska} topology. The \ac{LUSD} metric provides a notable enhancement to this measurement approach. While based on similar concepts, \ac{LUSD} offers a more comprehensive evaluation, considering both underutilization and overutilization.

All metrics presented in this study are derived from link utilization measurements. Link utilization values can be analyzed through multiple methodologies to characterize network load distribution resulting from processed traffic. The analytical approaches demonstrated in this study include statistical percentile calculations and threshold exceedance analysis of the collected utilization data. These can be further processed using either absolute difference calculations or Shapley value analysis to quantify individual flow contributions to network state. The key aspect of the described metrics is the selection of the $\alpha$ parameter, which determines metric sensitivity and influences the detection threshold for network state changes. Despite challenges in interpreting Shapley value-based metrics, incorporating them into the analysis enhances the understanding of flow impact. In practical applications, combining complementary metrics, particularly \ac{LUPD}, \ac{LUSD}, and \ac{TLUSSSV}, may be necessary for comprehensive assessment. These three metrics deliver maximum insight into network state while maintaining relatively low computational complexity. Additional insights can also be provided by \ac{RGPD} and \ac{TRGSSD}. Collectively, these metrics enable effective traffic engineering and support network function development by providing assessment of how proposed mechanisms affect overall network state.

\subsection{Practical applications}

To demonstrate the practical utility of our proposed flow impact metrics, we present two implementation scenarios. The first scenario addresses the detection and management of elephant flows to prevent network congestion and enhance reliability. The second scenario employs these metrics to evaluate various traffic management mechanisms, facilitating the selection of optimal solutions for network performance enhancement. Both cases illustrate how these metrics can inform operational decisions regarding flow management, design choices, and system configuration parameters.

\subsubsection{Detecting and handling elephant flows}

Elephant flows, characterized by large and persistent data transfers, can dominate link utilization and lead to network congestion~\cite{guo:diff}. Flow impact metrics enable early identification of these flows, allowing proactive management through routing mechanisms to reduce resource starvation risks.

Flow impact metrics quantify each flow's contribution to network load, identifying potential congestion sources. Our experimental evaluation revealed that \ac{LUPD} values correlated with bandwidth increases, consistently rising by approximately $0.20$ per bandwidth increment (Fig.~\ref{fig:lupd-a010}). Under high load conditions, \ac{LUPD} values reached $0.70$, confirming significant congestion contribution from the examined flows. The \emph{polska} topology analysis further demonstrated this pattern, where high load scenarios (H batches) produced \ac{TLUSSD} values exceeding $0.70$, indicating frequent congestion events (Fig.~\ref{fig:tlussd-a010}). 

Furthermore, the \ac{TLUSSSV} metric can pinpoint certain flows by considering both initial background load and introduced traffic. In the \emph{abilene} topology with H batches (Fig.~\ref{fig:tlussd-tlusssv-a010}), \ac{TLUSSSV} exceeded $0.50$, establishing the examined flow as a primary contributor to link saturation. As presented in Fig.~\ref{fig:tlussd-lusd-a020}, \ac{LUSD} occasionally reported slightly negative values when evaluating initially underutilized networks. Conversely, strongly positive \ac{LUSD} values exceeding $0.40$, as observed with the \emph{polska-H} batches, identify flows that precipitate saturation and should be flagged for mitigation.

In this scenario, flows exceeding critical threshold metric values (e.g., \ac{LUPD} or \ac{TLUSSD} > $0.70$) are classified as elephant flows. Once identified, these flows can be rerouted, rate-limited, or load balanced using \ac{SDN}-based mechanisms to preserve network performance.

\subsubsection{Evaluating network applications}

Network mechanisms produce distinctive traffic patterns affecting overall load distribution. Bursty traffic --- recurring high-volume transmission spikes --- can cause transient link overutilization and degrade performance for latency-sensitive applications. Such patterns are observed in measurement mechanisms implementing the OpenFlow protocol~\cite{rzepka:sfsa}. Flow impact metrics provide a systematic framework for evaluating these traffic characteristics and precisely identifying mechanisms that generate undesirable traffic burst patterns.

Flow impact metrics enable quantification of each application's contribution to network load profiles. Our analysis demonstrated that applications generating bursty traffic increase peak utilization measurements, with \ac{TLUSSD} values rising sharply in scenarios dominated by sudden transmission bursts (Fig.~\ref{fig:tlussd-a010}). Interestingly, we identified multiple instances where \ac{LUSD} registered slightly negative values (approx. $-0.10$) in moderate-load scenarios (e.g., \emph{abilene-M} with $\alpha=0.10$). These values indicate that adding new flows enhances resource utilization efficiency without approaching capacity limitations. Conversely, strongly positive \ac{LUSD} values (e.g., exceeding $0.60$) indicate that the examined flows are driving network utilization beyond sustainable thresholds, requiring immediate mitigation.

Comparative analysis of metric values enables identification of mechanisms generating problematic traffic patterns. This evaluation methodology improves selection of network management solutions that optimize resource allocation while minimizing congestion. For instance, when an OpenFlow-based measurement system~\cite{rzepka:sfsa} generates traffic bursts (e.g., indicated by \ac{TLUSSD} or \ac{LUSD} values exceeding $0.60$), administrators can implement optimizations by adjusting polling parameters to maintain metrics below established thresholds. These insights directly guide design choices and configuration parameter selection during network planning.

\section{Conclusions}
\label{sec:conclusions}

This paper presents a comprehensive comparative analysis of 11 metrics that quantify flow impact on network state. These metrics offer valuable tools for traffic engineering research and network mechanism development. We have detailed both the fundamental concepts underlying these metrics and the methodologies for their definition. Our original contributions include the Utilization Score metric and modifications to percentile calculation and Shapley Value application in network analysis. 

A primary research objective was to identify metrics that function independently of specific network mechanisms, accept input from any sufficient data source, and operate without requiring modifications to production networks. Most of the evaluated solutions successfully meet these criteria. Our assessment demonstrates that several metrics, despite their relatively straightforward implementation based on established concepts, effectively characterize network state modifications caused by specific flows. The selection of appropriate metrics should be tailored to particular use case requirements. Based on our experimental results, we recommend the combination of \ac{LUPD}, \ac{LUSD}, and \ac{TLUSSSV} metrics, which collectively provide comprehensive analytical coverage while maintaining reasonable implementation complexity.

The metric construction methodology presented in this paper establishes a framework for future research. This approach can be extended through additional analysis methods and foundational metrics to develop new techniques for assessing flow impact on network performance.

\printbibliography

\begin{IEEEbiographynophoto}{\MakeUppercase{Michał Rzepka}} holds a PhD degree in Telecommunications from AGH University, which he obtained in 2023, preceded by an MSc in ICT from the same university in 2017. His research activity revolves around Software Defined Networking, traffic engineering, and network measurements.
\end{IEEEbiographynophoto}

\begin{IEEEbiographynophoto}{\MakeUppercase{Piotr Chołda}} graduated from AGH University of Krakow in 2001. Afterwards, he obtained a doctorate in telecommunications in 2006 from the same university. Then, he joined the Institute of Telecommunications there, and now he works as an Associate Professor. He specializes in design and management of ICT systems.
\end{IEEEbiographynophoto}
\EOD
\end{document}